\documentclass{sig-alternate}
\usepackage{mathptmx}       
\usepackage{helvet}         
\usepackage{courier}        
\usepackage{type1cm}        
\usepackage{makeidx}         
\usepackage{graphicx}        
\usepackage{multicol}        
\usepackage[bottom]{footmisc}
\usepackage{graphicx}
\usepackage{latexsym}
\usepackage{amsmath}
\usepackage{amssymb}
\usepackage{fancyvrb}
\usepackage{subfigure}
\usepackage{colortbl}
\usepackage{enumerate}
\usepackage{pifont}
\usepackage{stmaryrd}
\usepackage{textcomp}
\usepackage{fncylab}
\usepackage{multirow}
\usepackage{paralist}
\usepackage{wrapfig}
\usepackage{colortbl}
\graphicspath {{./} {Figures/}}

\newcolumntype{H}{>{\columncolor{black}\color{white}}c}

\usepackage[ruled]{algorithm2e}

\SetAlFnt{\small}
\SetAlCapFnt{\small}
\SetAlCapNameFnt{\small}
\SetAlCapHSkip{0pt}
\IncMargin{-\parindent}

\begin{document}

\title{The Family of MapReduce and Large Scale Data Processing Systems}
\numberofauthors{3}
\author{
%
\alignauthor Sherif Sakr\\
       \affaddr{NICTA and University of New South Wales}\\
       \affaddr{Sydney, Australia}\\
       \email{ssakr@cse.unsw.edu.au}
\alignauthor Anna Liu\\
       \affaddr{NICTA and University of New South Wales}\\
       \affaddr{Sydney, Australia}\\
       \email{anna.liu@nicta.com.au}
\alignauthor Ayman G. Fayoumi,\\
       \affaddr{King Abdulaziz University}\\
       \affaddr{Jeddah, Saudia Arabia}\\
       \email{afayoumi@kau.edu.sa}
}

\maketitle

\begin{abstract}
In the last two decades, the continuous increase of computational power has produced an overwhelming flow of data which has called for a paradigm shift in the computing architecture and large scale data processing mechanisms. MapReduce is a simple and powerful programming model that enables easy development of scalable parallel applications to process vast amounts of data on large clusters of commodity machines.  It isolates the application from the details of running a distributed program such as issues on data distribution, scheduling and fault tolerance. However, the original implementation of the MapReduce framework had some limitations that have been tackled by many research efforts in several followup works after its introduction. This article provides a comprehensive survey for a \emph{family} of  approaches and mechanisms of large scale data processing mechanisms that have been implemented based on the original  idea of the MapReduce framework and are currently gaining a lot of momentum in both research and industrial communities. We also cover a set of introduced systems that have been implemented to provide declarative programming interfaces on top of the MapReduce framework. 
In addition, we review several large scale data processing systems that resemble some of the ideas of the MapReduce framework for different purposes and application scenarios.
Finally, we discuss some of the future research directions for implementing the next generation of
MapReduce-like solutions.
\end{abstract}

\section{Introduction}
We live in the era of \emph{Big Data} where we are witnessing a continuous increase on the computational power that produces an overwhelming flow of data which has called for a \emph{paradigm shift} in the computing architecture and large scale data processing mechanisms.  Powerful telescopes in astronomy, particle accelerators in physics, and genome sequencers in biology are putting massive volumes of data into the hands of scientists. For example, the Large Synoptic Survey Telescope~\cite{LSST} generates on the order of 30 TeraBytes of data every day.
Many enterprises continuously collect large datasets that record customer interactions, product sales, results from advertising campaigns on the Web, and other types of information. For example, Facebook collects 15 TeraBytes of data each day into a PetaByte-scale data warehouse~\cite{Hive1}. Jim Gray, called the shift a "\emph{fourth paradigm}"~\cite{FourthParadigm}. The first three paradigms were \emph{experimental}, \emph{theoretical} and, more recently, \emph{computational science}. Gray argued that the only way to cope with this paradigm is to develop a new generation of computing tools to manage, visualize and analyze the data flood. In general, current computer architectures are increasingly imbalanced where the latency gap between multi-core CPUs and mechanical hard disks is growing every year which makes the challenges of data-intensive computing much harder to overcome~\cite{Petascale}. Hence, there is a crucial need for a systematic and generic approach to tackle these problems with an architecture that can also scale into the foreseeable future~\cite{Sakr11}. In response, Gray argued that the new trend should instead focus on supporting cheaper clusters of computers to manage and process all this data instead of focusing on having the biggest and fastest single computer.

In general, the growing demand for large-scale data mining and data analysis applications has spurred the development of novel solutions from both the industry (e.g., web-data analysis, click-stream analysis, network-monitoring log analysis) and the sciences (e.g., analysis of data produced by massive-scale simulations, sensor deployments, high-throughput lab equipment). Although parallel database systems~\cite{PDB}  serve some of these data analysis applications (e.g. Teradata\footnote{http://teradata.com/}, SQL Server PDW\footnote{http://www.microsoft.com/sqlserver/en/us/solutions-technologies/data-warehousing/pdw.aspx},
Vertica\footnote{http://www.vertica.com/},
Greenplum\footnote{http://www.greenplum.com/},
ParAccel\footnote{http://www.paraccel.com/},
Netezza\footnote{http://www-01.ibm.com/software/data/netezza/}), they are expensive, difficult to administer and lack fault-tolerance for long-running queries~\cite{HadoopDB2}.  MapReduce~\cite{MapReduce1} is a framework which is introduced by Google for programming commodity computer clusters to perform large-scale data processing in a single pass. The framework is designed such that a MapReduce cluster can scale to thousands of nodes in a fault-tolerant manner. One of the main advantages of this framework is its reliance on a simple and powerful programming model. In addition, it isolates the application developer from all the complex details of running a distributed program such as: issues on data distribution, scheduling and fault tolerance~\cite{DataCenter}.

Recently, there has been a great deal of hype about cloud computing~\cite{BerkelyCloud}. In principle, cloud computing is associated with a new paradigm for the provision of computing infrastructure. This paradigm shifts the location of this infrastructure to more centralized and larger scale datacenters in order to reduce the costs associated with the management of hardware and software resources. In particular, cloud computing  has promised a number of advantages for hosting the deployments of data-intensive applications such as:
\begin{compactitem}
\item Reduced time-to-market by removing or simplifying the time-consuming hardware provisioning, purchasing and deployment processes.
\item Reduced monetary cost by following a \emph{pay-as-you-go} business model.
\item Unlimited (virtually) throughput by adding servers if the workload increases.
\end{compactitem}
In principle, the success of many enterprises often rely on their ability to analyze expansive volumes of data.  In general, cost-effective processing of large datasets is a nontrivial undertaking. Fortunately, MapReduce frameworks and cloud computing have made it easier than ever for everyone to step into the world of big data. This technology combination has enabled even small companies to collect and analyze terabytes of data in order to gain a competitive edge. For example, the Amazon  Elastic Compute Cloud (EC2)\footnote{http://aws.amazon.com/ec2/} is offered as a commodity that can be purchased and utilised. In addition, Amazon has also provided the Amazon Elastic MapReduce\footnote{http://aws.amazon.com/elasticmapreduce/} as an online service to easily and cost-effectively process vast amounts of data without the need to worry about time-consuming set-up, management or tuning of computing clusters or the compute capacity upon which they sit. Hence, such services enable third-parties to perform their analytical queries on massive datasets with minimum effort and cost by abstracting the complexity entailed in building and maintaining computer clusters.

The implementation of the basic MapReduce architecture had some limitations. Therefore, several research efforts have been triggered to tackle these limitations by introducing several advancements in the basic architecture in order to improve its performance. This article provides a comprehensive survey for a \emph{family} of approaches and mechanisms of large scale data analysis mechanisms that have been implemented based on the original  idea of the MapReduce framework and are currently gaining a lot of momentum in both research and industrial communities. In particular, the remainder of this article is organized as follows. Section \ref{SEC:MapReduce} describes the basic architecture of the MapReduce framework. Section \ref{SEC:MapReduceImprove} discusses several techniques that have been proposed to improve the performance and capabilities of the MapReduce framework from different perspectives. Section~\ref{SEC:SQLLike} covers several systems that support a high level SQL-like interface for the MapReduce framework. 
Section~\ref{SEC:SYSTEMS} reviews several large scale data processing systems that resemble some of the ideas of the MapReduce framework, without sharing its architecture or infrastructure,  for different purposes and application scenarios. In Section~\ref{SEC:Conclusions}, we conclude the article  and discuss some of the future research directions for implementing the next generation of MapReduce/Hadoop-like solutions.

\section{MapReduce Framework: Basic Architecture}
\label{SEC:MapReduce}

The MapReduce framework is introduced as a simple and powerful programming model that enables easy development of scalable parallel applications to process vast amounts of data on large clusters of commodity machines~\cite{MapReduce1,MapReduce2}. In particular, the implementation described in the original paper is mainly designed to achieve high performance on large clusters of commodity PCs. One of the main advantages of this approach is that it isolates the application from the details of running a distributed program, such as issues on data distribution, scheduling and fault tolerance. In this model, the computation takes a set of key/value pairs input and produces a set of  key/value pairs as output. The user of the MapReduce framework expresses the computation using two functions: \emph{Map} and \emph{Reduce}. The Map function takes an input pair and produces a set of intermediate key/value pairs. The MapReduce framework groups together all intermediate values associated with the same intermediate key $I$ and passes them to the Reduce function. The Reduce function receives an intermediate key $I$ with its set of values and merges them together. Typically just zero or one output value is produced per Reduce invocation. The main advantage of this model is that it allows large computations to be easily parallelized and re-executed to be used as the primary mechanism for fault tolerance. Figure \ref{Fig:MR1} illustrates an example MapReduce program expressed in pseudo-code for counting the number of occurrences of each word in a collection of documents. In this example, the map function emits each word plus an associated count of occurrences while the reduce function sums together all counts emitted for a particular word. In principle, the design of the MapReduce framework has considered the following  main principles~\cite{MapReduce5}:
\begin{figure}[t]
  \centering
  \includegraphics[width=0.5\textwidth]{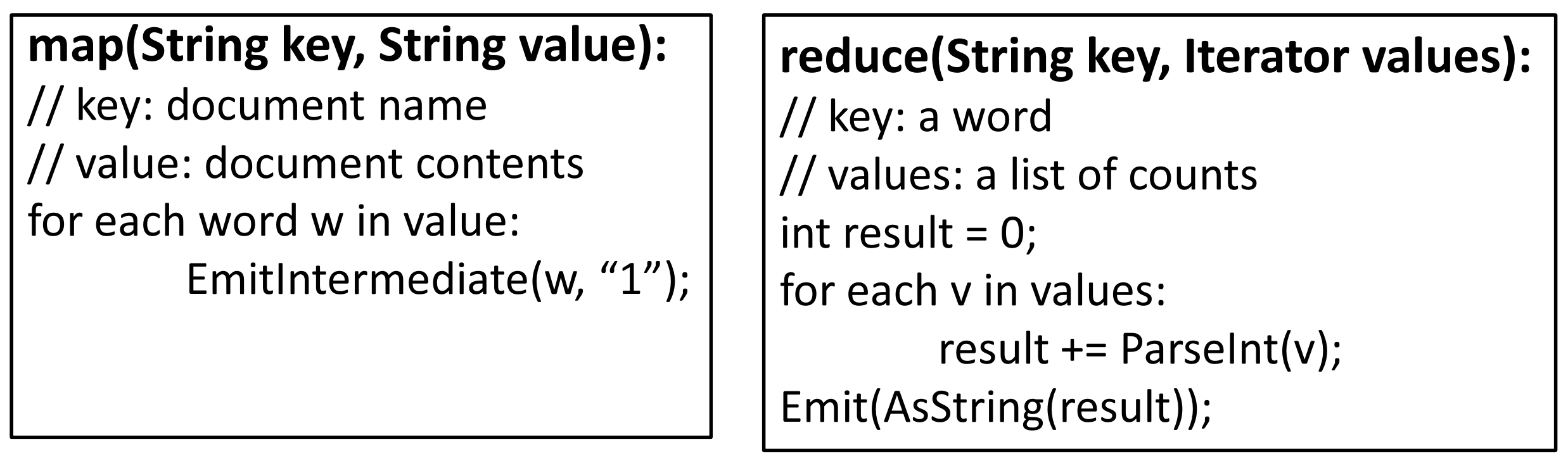}\\
  \caption{An Example MapReduce Program~\cite{MapReduce1}.}\label{Fig:MR1}
\end{figure}

\begin{compactitem}
\item \emph{Low-Cost Unreliable Commodity Hardware}: Instead of using expensive, high-performance,  reliable symmetric multiprocessing (SMP) or massively parallel processing (MPP) machines equipped with high-end network and storage subsystems, the MapReduce framework  is designed to run on large clusters of commodity hardware. This hardware is managed and powered by open-source operating systems and utilities so that the cost is low.
\item \emph{Extremely Scalable RAIN Cluster}: Instead of using centralized RAID-based SAN or NAS storage systems, every MapReduce node has its own local off-the-shelf hard drives. These nodes are loosely coupled where they are placed in racks that can be connected with standard networking hardware connections. These nodes can be taken out of service with almost no impact to still-running MapReduce jobs. These clusters are called Redundant Array of Independent (and Inexpensive) Nodes (RAIN).
\item \emph{Fault-Tolerant yet Easy to Administer}: MapReduce jobs can run on clusters with thousands of nodes or even more. These nodes are not very reliable as at any point in time, a certain percentage of these commodity nodes or hard drives will be out of order. Hence, the MapReduce framework applies straightforward mechanisms to replicate data and launch backup tasks so as to keep still-running processes going. To handle crashed nodes, system administrators simply take crashed hardware off-line. New nodes can be plugged in at any time without much administrative hassle. There is no complicated backup, restore and recovery configurations like the ones that can be seen in many DBMSs.
\item \emph{Highly Parallel yet Abstracted}: The most important contribution of the MapReduce framework is its ability to automatically support the parallelization of task executions. Hence, it allows developers to focus mainly on the problem at hand rather than worrying about the low level implementation details such as memory management, file allocation, parallel, multi-threaded or network programming. Moreover, MapReduce's shared-nothing architecture~\cite{SharedNothing} makes it much more scalable and ready for parallelization.

\end{compactitem}
\begin{figure}[t]
  \centering
  \includegraphics[width=0.5\textwidth]{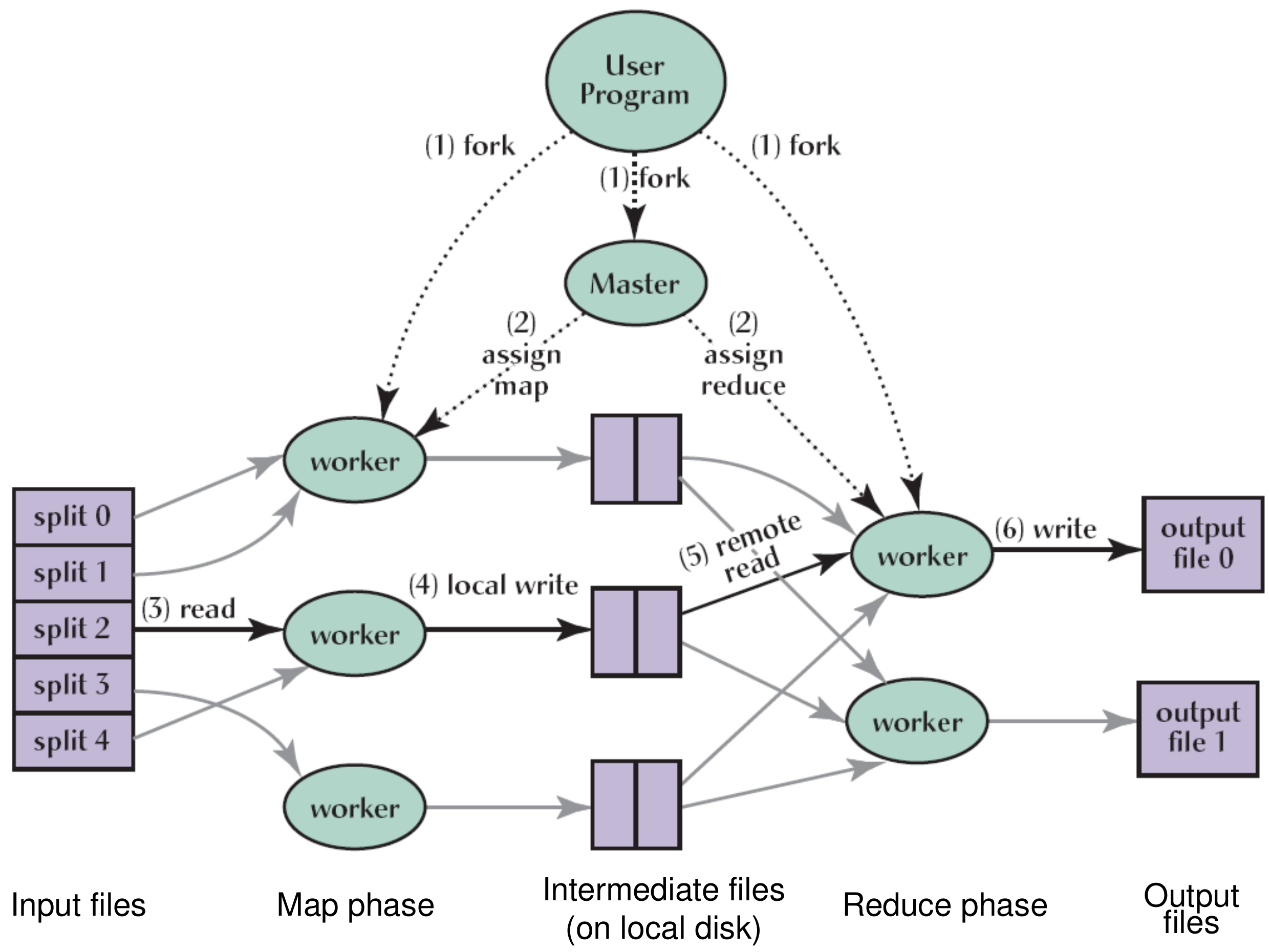}\\
  \caption{An Overview of the Flow of Execution a MapReduce Operation~\cite{MapReduce1}.}\label{Fig:MR2}
\end{figure}

Hadoop\footnote{http://hadoop.apache.org/} is an open source Java library~\cite{HadoopGuide} that supports data-intensive distributed applications by realizing the implementation of the MapReduce framework\footnote{In the rest of this article, we use the two names: MapReduce and Hadoop, interchangeably}. It has been widely used by a large number of business companies for production purposes\footnote{http://wiki.apache.org/hadoop/PoweredBy}. On the implementation level, the Map invocations of a MapReduce job are distributed across multiple machines by automatically partitioning the input data into a set of $M$ splits. The input splits can be processed in parallel by different machines. Reduce invocations are distributed by partitioning the intermediate key space into $R$ pieces using a partitioning function (e.g. hash(key) mod R). The number of partitions (R) and the partitioning function are specified by the user. Figure~\ref{Fig:MR2} illustrates an example of the overall flow of a MapReduce operation which goes through the following sequence of actions:

\begin{compactenum}
 \item The input data of the MapReduce program is split into $M$ pieces and starts up many instances of the program on a cluster of machines.
 \item  One of the instances of the program is elected to be the \emph{master} copy while the rest are considered as \emph{workers} that are assigned their work by the master copy. In particular, there are $M$ map tasks and $R$ reduce tasks to assign. The master picks idle workers and assigns each one or more map tasks and/or reduce tasks.
 \item  A worker who is assigned a map task processes the contents of the corresponding input split and generates key/value pairs from the input data and passes each pair to the user-defined Map function. The intermediate key/value pairs produced by the Map function are buffered in memory.
 \item  Periodically, the buffered pairs are written to local disk and partitioned into $R$ regions by the partitioning function. The locations of these buffered pairs on the local disk are passed back to the master, who is responsible for forwarding these locations to the reduce workers.
 \item  When a reduce worker is notified by the master about these locations, it reads the buffered data from the local disks of the map workers which is then sorted by the intermediate keys so that all occurrences of the same key are grouped together. The sorting operation is needed because typically many different keys map to the same reduce task.
 \item  The reduce worker  passes the key and the corresponding set of intermediate values to the user's Reduce function. The output of the Reduce function is appended to a final output file for this reduce partition.
  \item When all map tasks and reduce tasks have been completed, the master program wakes up the user program. At this point, the MapReduce invocation in the user program returns the program control back to the user code.
\end{compactenum}

During the execution process, the master pings every worker periodically. If no response is received from a worker within a certain amount of time, the master marks the worker as \emph{failed}. Any map tasks marked \emph{completed} or \emph{in progress} by the worker are reset back to their initial idle state and therefore become eligible for scheduling by other workers. Completed map tasks are re-executed on a task failure because their output is stored on the local disk(s) of the failed machine and is therefore inaccessible. Completed reduce tasks do not need to be re-executed since their output is stored in a global file system.

\section{Extensions and Enhancements of The MapReduce Framework}
\label{SEC:MapReduceImprove}

In practice, the basic implementation of the MapReduce is very useful for handling data processing and data loading in a heterogenous system with many different storage systems. Moreover, it provides a flexible framework for the execution of more complicated functions than that can be directly supported  in SQL. However, this basic architecture suffers from some limitations. ~\cite{MapReduce3} reported about some possible improvements that can be incorporated into the MapReduce framework. Examples of these possible improvements include:
\begin{compactitem}
\item MapReduce  should take advantage of natural indices  whenever possible.
\item Most MapReduce output can be left unmerged since there is no benefit of merging them if the next consumer is just  another MapReduce program.
\item MapReduce users should avoid using inefficient textual formats.
\end{compactitem}

In the following subsections we discuss some research efforts that have been conducted in order to deal with these challenges and the different improvements that has been made on the basic implementation of the  MapReduce framework in order to achieve these goals.

\begin{figure}[t]
  \centering
  \includegraphics[width=0.5\textwidth]{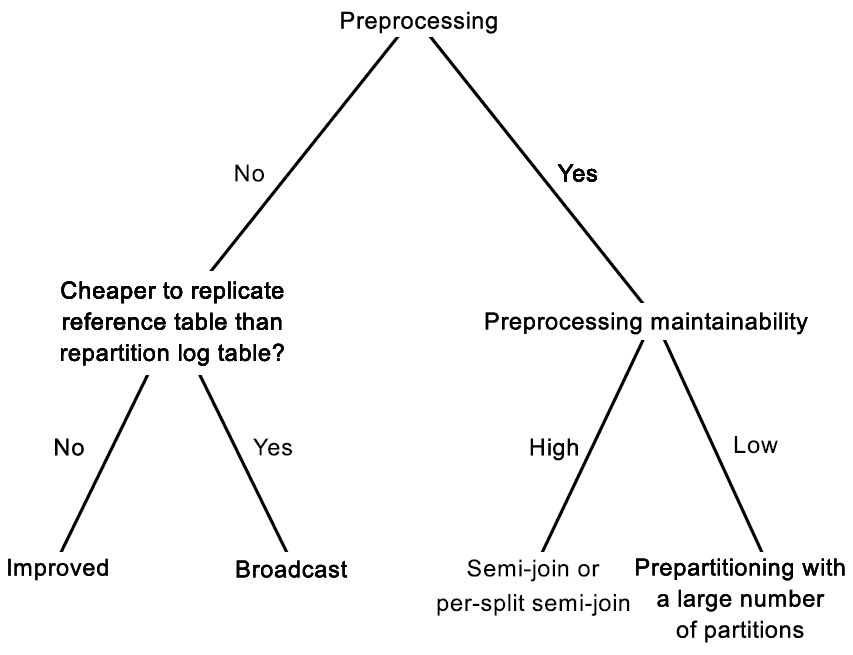}\\
  \caption{Decision tree for choosing between various join strategies on the MapReduce Framework~\cite{JoinComparison}.}\label{Fig:JC}
\end{figure}

\subsection{Processing Join Operations}
One main limitation of the MapReduce framework is that it does not support the joining of multiple datasets in one task. However, this can still  be achieved with additional MapReduce steps. For example, users can map and reduce one dataset and read data from other datasets on the fly. Blanas et al.~\cite{JoinComparison} have reported about a study that evaluated the performance of different distributed join algorithms using the MapReduce framework. In particular, they have evaluated the following implementation strategies of distributed join algorithms:

\begin{compactitem}
\item \emph{Standard repartition join}: The two input relations are dynamically partitioned on the join key and the corresponding pairs of partitions are joined using the standard partitioned sort-merge join approach.
\item \emph{Improved repartition join}: One potential problem with the standard repartition join is that all the records for a given join key from both input relations have to be buffered. Therefore, when the key cardinality is small or when the data is highly skewed, all the records for a given join key may not fit in memory. The improved repartition join strategy fixes the buffering problem by introducing the following key changes:

\begin{compactitem}
\item In the map function, the output key is changed to a composite of the join key and the table tag. The table tags are generated
in a way that ensures records from one input relation will be sorted ahead of those from the other input relation on a given join key.
\item  The partitioning function is customized so that the hashcode is computed from just the join key part of the composite key. This way records with the same join key are still assigned to the same reduce task.
\item As records from the smaller input are guaranteed to be ahead of those from L for a given join key, only the records from the smaller input are buffered and the records of the larger input are streamed to generate the join output.
\end{compactitem}
\item \emph{Broadcast join}: Instead of moving both input relations across the network as in the repartition-based
joins, the broadcast join approach moves only the smaller input relation so that it  avoids the preprocessing sorting requirement of  both input relations and more importantly avoids the network overhead for moving the larger relation.
\item \emph{Semi-join}: This join approach tries to avoid the problem of the broadcast join approach where it is possible to send many records of the smaller input relation across the network while they may not be actually referenced by any records in the other relation.
    It achieves this goal at the cost of an extra scan of the smaller input relation where it determines the set of unique join keys in the smaller relation, send them to the other relation to specify the list of the actual referenced join keys and then send only these records across the network for executing the real execution of the join operation.
\item \emph{Per-split semi-join}: This join approach tries to improve the semi-join approach with a further step to address the fact that not every record in the filtered version of the smaller relation will join with a particular split of the larger relation. Therefore, an extra process step is executed to determine the target split(s) of each filtered join key.
\end{compactitem}

Figure \ref{Fig:JC} illustrates a decision tree that summarizes the tradeoffs of the studied join strategies according to the results of that study. Based on statistics, such as the relative data size and the fraction of the join key referenced, this decision tree tries to determine what is the right join strategy for a given circumstance. If data is not preprocessed, the right join strategy depends on the size of the data transferred via the network. If the network cost of broadcasting an input relation $R$ to every node is less expensive than transferring both $R$ and projected $L$, then the broadcast join algorithm should be used. When preprocessing is allowed, semi-join, per-split semi-join and directed join with sufficient partitions are the best choices. Semi-join and per-split semi-join offer further flexibility since their preprocessing steps are insensitive to how the log table is organized, and thus suitable for any number of reference tables. In addition, the preprocessing steps of these two algorithms are cheaper since there is no shuffling of the log data.

\begin{figure}[t]
  \centering
  \includegraphics[width=0.5\textwidth]{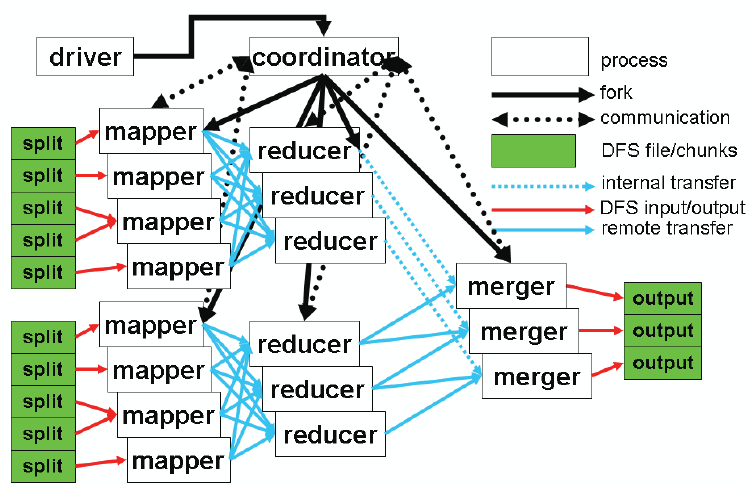}\\
  \caption{An Overview of The Map-Reduce-Merge Framework~\cite{MapReduce5}.}\label{Fig:MRM}
\end{figure}

\begin{figure}[t]
  \centering
  \includegraphics[width=0.5\textwidth]{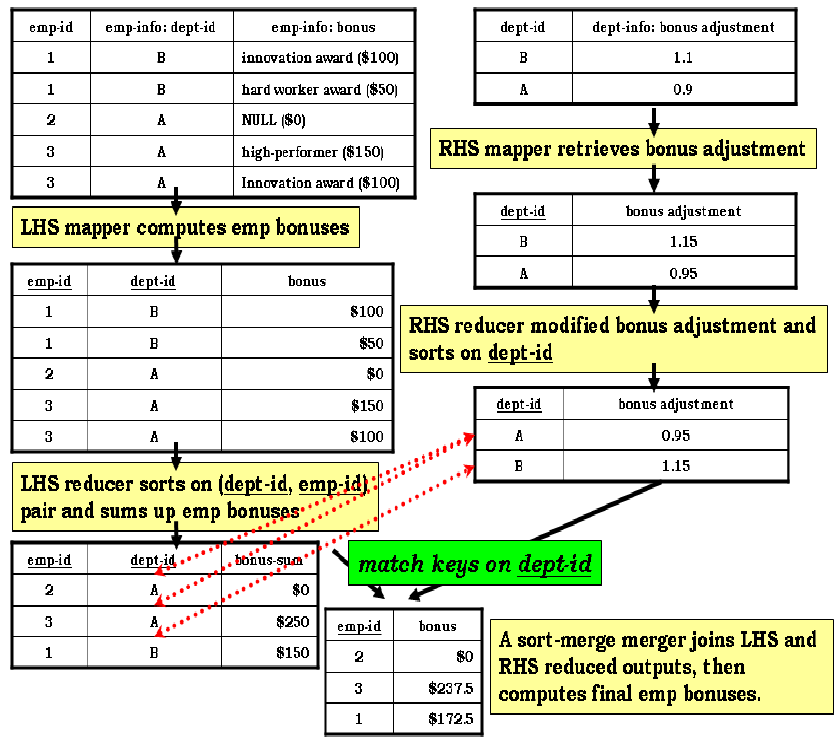}\\
  \caption{A Sample Execution of The Map-Reduce-Merge Framework~\cite{MapReduce5}.}\label{Fig:MRM2}
\end{figure}

To tackle the limitation of the extra processing requirements for performing join operations in the MapReduce framework, the \emph{Map-Reduce-Merge} model~\cite{MapReduce5}  have been introduced to  enable the processing of multiple  datasets. Figure \ref{Fig:MRM} illustrates the framework of this model where the map phase transforms an input key/value pair $(k1, v1)$ into a list of intermediate key/value pairs $[(k2, v2)]$. The reduce function aggregates the list of values $[v2]$ associated with $k2$ and produces a list of values $[v3]$ which is also associated with k2. Note that inputs and outputs of both functions belong to the same lineage ($\alpha$). Another pair of map and reduce functions produce the intermediate output $(k3, [v4])$ from another lineage ($\beta$). Based on keys $k2$ and $k3$, the merge function combines the two reduced outputs from different lineages into a list of key/value outputs $[(k4, v5)]$. This final output becomes a new lineage ($\gamma$). If $\alpha$ = $\beta$ then this merge function does a self-merge which is similar to self-join in relational algebra. The main differences between the processing model of this framework and the original MapReduce is the production of a key/value list from the reduce function instead of just that of values. This change is introduced because the merge function requires input datasets to be organized (partitioned, then either sorted or hashed) by keys and these keys have to be passed into the function to be merged. In the original framework, the reduced output is final. Hence, users pack whatever is needed in $[v3]$ while passing $k2$ for the next stage is not required. Figure~\ref{Fig:MRM2} illustrates a sample execution of the Map-Reduce-Merge framework. In this example, there are two datasets \emph{Employee} and \emph{Department} where Employee's key attribute is \texttt{emp-id} and the Department's key is \texttt{dept-id}. The execution of this example query aims to join these two datasets and compute employee bonuses. On the left hand side of Figure~\ref{Fig:MRM2}, a mapper reads Employee entries and computes a bonus for each entry. A reducer then sums up these bonuses for every employee and sorts them by \texttt{dept-id}, then \texttt{emp-id}. On the right hand side, a mapper reads Department entries and computes bonus adjustments.
A reducer then sorts these department entries.  At the end, a merger matches the output records from the two reducers on \texttt{dept-id} and applies a department-based bonus adjustment on employee bonuses. Yang et al.~\cite{MapReduce6} have also proposed an approach for improving the Map-Reduce-Merge framework by adding a new primitive called \emph{Traverse}. This primitive can process index file entries recursively, select data partitions based on query conditions and feed only selected partitions to other primitives.

The \emph{Map-Join-Reduce}~\cite{MJR} represents another approach that has been introduced with a filtering-join-aggregation programming
model as an extension of the standard MapReduce's filtering-aggregation programming model. In particular, in addition to the standard mapper and reducer operation of the standard MapReduce framework, they introduce a third operation, join (called joiner), to the framework. Hence, to join multiple datasets for aggregation, users specify a set of \emph{join}() functions and the join order between them. Then, the runtime system automatically joins the multiple input datasets according to the join order and invoke \emph{join}() functions to process the joined records. They have also introduced a one-to-many shuffling strategy which shuffles each intermediate key/value pair to many joiners at one time. Using a tailored partition strategy, they can utilize the one-to-many shuffling scheme to join multiple datasets in one phase instead of a sequence of MapReduce jobs.  The runtime system for executing a Map-Join-Reduce job  launches two kinds of processes: \emph{MapTask}, and \emph{ReduceTask}. Mappers run inside the MapTask process while joiners and reducers are invoked inside the ReduceTask process. Therefore, Map-Join-Reduce's process model allows for the pipelining of intermediate results between joiners and reducers since joiners and reducers are run inside the same ReduceTask process.

~\cite{MRJoin,MultiwayJoin} have presented another approach to improve the join phase in the MapReduce framework.
The approach aims to optimize the communication cost by focusing on selecting the most appropriate attributes that are used to partition and replicate the data among the reduce process.
Therefore, it begins by identifying the \emph{map-key} the set of attributes that identify the Reduce process to which a Map process must send a particular tuple. Each attribute of the map-key gets a "\emph{share}" which is the number of buckets into which its values are hashed, to form a component of the identifier of a Reduce process. Relations have their tuples replicated in limited fashion of which the degree of replication depends on the shares for those map-key attributes that are missing from their schema.
The approach consider two important special join cases: \emph{chain} joins (represents a sequence of 2-way join operations where the output of one operation in this sequence is used as an input to another operation in a pipelined fashion) and \emph{star} joins (represents joining of a large fact table with several smaller dimension tables). In each case, the proposed algorithm is able to determine the map-key and determine the shares that yield the least replication. The proposed approach is not always superior to the conventional way of using map-reduce to implement joins. However, there are some cases where the proposed approach results in clear wins such as: 1) Analytic queries in which a very large fact table is joined with smaller dimension tables. 2) Queries involving paths through graphs with high out-degree, such as the Web or a social network.

\subsection{Supporting Iterative Processing}

The basic MapReduce framework does not directly support these iterative data analysis applications. Instead, programmers must implement iterative programs by manually issuing multiple MapReduce jobs and orchestrating their execution using a driver program. In practice, there are two key problems with manually orchestrating an iterative program in MapReduce:
\begin{compactitem}
\item Even though much of the data may be unchanged from iteration to iteration, the data must be re-loaded and re-processed at each iteration, wasting I/O, network bandwidth and CPU resources.
\item The termination condition may involve the detection of when a fixpoint has been reached. This condition may itself require an extra MapReduce job on each iteration, again incurring overhead in terms of scheduling extra tasks, reading extra data from disk and moving data across the network.
\end{compactitem}

 The  \emph{HaLoop} system~\cite{HaLoop} is designed to support iterative processing on the MapReduce framework by extending the basic MapReduce framework with two main functionalities:
\begin{compactenum}
\item Caching the invariant data in the first iteration and then reusing them in later iterations.
\item Caching the reducer outputs, which makes checking for a fixpoint more efficient, without an extra MapReduce job.
\end{compactenum}

In order to accommodate the requirements of iterative data analysis applications, HaLoop has incorporated the following changes to the basic Hadoop MapReduce framework:
\begin{compactitem}
\item It exposes a new application programming interface to users that simplifies the expression of iterative MapReduce programs.
\item HaLoop's master node contains a new loop control module that repeatedly starts new map-reduce steps that compose the loop body until a user-specified stopping condition is met.
\item It uses a new task scheduler that leverages data locality.
\item It caches and indices application data on slave nodes. In principle, the task tracker not only manages task execution but also manages caches and indices on the slave node and redirects each task's cache and index accesses to local file system.
\end{compactitem}
In principle, HaLoop relies on the same file system and has the same task queue structure as Hadoop but the task scheduler and task tracker modules are modified, and the loop control, caching, and indexing modules are newly introduced to the architecture. The task tracker not only manages task execution but also manages caches and indices on the slave node, and redirects each task's cache and index accesses to local file system.

In the MapReduce framework, each map or reduce task contains its portion of the input data and
the task runs by performing the map/reduce function on its input data records where the life cycle of the task ends when finishing the processing of all the input data records has been completed. The \emph{iMapReduce} framework~\cite{iMapReduce} supports the feature of iterative processing by keeping alive each map and reduce task  during the whole iterative process. In particular, when all of the input data of a persistent task are parsed and processed, the task becomes dormant, waiting for the new updated input data. For a map task, it waits for the results from the reduce tasks and is activated to work on the new input records when the required data from the reduce tasks arrive. For the reduce tasks, they wait for the map tasks' output and are activated synchronously as in MapReduce.
Jobs can terminate their iterative process in one of two ways:
\begin{compactenum}
\item \emph{Defining fixed number of iterations}: Iterative algorithm stops after it iterates \emph{n} times.
\item \emph{Bounding the distance between two consecutive iterations}: Iterative algorithm stops when the distance is less than a threshold.
\end{compactenum}
The iMapReduce runtime system does the termination check after each iteration.
To terminate the iterations by a fixed number of iterations, the persistent map/reduce task records its iteration number and terminates itself when the number exceeds a threshold.
To bound the distance between the output from two consecutive iterations, the reduce tasks can save the output from two
consecutive iterations and compute the distance.  If the termination condition is satisfied, the master will notify all the map and reduce tasks to terminate their execution.

Other projects have been implemented for supporting iterative processing on the MapReduce framework. For example, \emph{Twister}\footnote{http://www.iterativemapreduce.org/}  is a MapReduce runtime with an extended programming model that supports iterative MapReduce computations efficiently~\cite{Twister}. It uses a publish/subscribe messaging infrastructure for communication and data transfers, and supports long running map/reduce tasks. In particular, it provides programming extensions to MapReduce with broadcast and scatter type data transfers. Microsoft has also developed a project that provides an iterative MapReduce runtime for Windows Azure called \emph{Daytona}\footnote{http://research.microsoft.com/en-us/projects/daytona/}.

\subsection{Data and Process Sharing}
With the emergence of cloud computing, the use of an analytical query processing infrastructure (e.g., Amazon EC2) can be directly mapped to \emph{monetary} value. Taking into account that different MapReduce jobs can perform similar work, there could be many opportunities for sharing the execution of their work. Thus, this sharing can reduce the overall amount of work which consequently leads to the reduction of the monetary charges incurred while utilizing the resources of the processing infrastructure. The \emph{MRShare} system~\cite{MRShare} have been presented as a sharing framework which is tailored to  transform a batch of queries into a new batch that will be executed more efficiently by merging jobs into groups and evaluating each group as a single query. Based on a defined cost model, they described an optimization problem that aims to derive the optimal grouping of queries in order to avoid performing redundant work and thus resulting in significant savings on both processing time and money. In particular, the approach considers exploiting the following sharing opportunities:

\begin{compactitem}
\item \emph{Sharing Scans}. To share scans between two mapping pipelines $M_i$ and $M_j$, the input data must be the same.
In addition, the key/value pairs should be of the same type. Given that, it becomes possible to merge the two pipelines into a single pipeline and scan the input data only once. However, it should be noted that such combined mapping will produce two
streams of output tuples (one for each mapping pipeline $M_i$ and $M_j$) . In order to distinguish the streams
at the reducer stage, each tuple is tagged with a \texttt{tag()} part. This tagging part is used to indicate the origin mapping pipeline during the reduce phase.

\item \emph{Sharing Map Output}. If the map output key and value types are the same for two  mapping pipelines $M_i$ and $M_j$ then the map output streams for $M_i$ and $M_j$ can be shared. In particular, if $Map_i$ and $Map_j$ are applied to each input tuple. Then, the map output tuples coming only from $Map_i$ are tagged with \texttt{tag(i)} only. If a map output tuple was produced from an input tuple by both $Map_i$  and $Map_j$, it is then tagged by \texttt{tag(i)+tag(j)}. Therefore, any overlapping parts of the map output will be shared. In principle, producing a smaller map output leads to savings on sorting and copying intermediate data over the network.

\item \emph{Sharing Map Functions}. Sometimes the map functions are identical and thus they can be executed once. At the end
of the map stage two streams are produced, each tagged with its job tag. If the map output is shared, then clearly only one stream
needs to be generated. Even if only some filters are common in both jobs, it is possible to share parts of map functions.
\end{compactitem}

In practice, sharing scans and sharing map-output yield I/O savings while sharing map functions (or parts of them) additionally yield CPU savings.

While the \emph{MRShare} system focus on sharing the processing  between queries that are executed concurrently, the \emph{ReStore} system~\cite{ReStore2,ReStore} has been introduced so that it can enable the queries that are submitted at different times to share the intermediate results of previously executed jobs and reusing them for future submitted jobs to the system. In particular, each MapReduce job  produces output that is stored in the distributed file system used by the
MapReduce system (e.g. HDFS). These intermediate results are kept (for a defined period) and managed so that they can be used as input by subsequent jobs. ReStore can make use of whole jobs or sub-jobs reuse opportunities. To achieve this goal, the ReStore consists of two main components:
\begin{compactitem}
\item \emph{Repository of MapReduce job outputs}: It stores the outputs of previously executed MapReduce jobs and the physical plans of these jobs.
\item \emph{Plan matcher and rewriter}: Its aim is to find physical plans in the repository that can be used to rewrite the input jobs using the available matching intermediate results.
\end{compactitem}
In principle, the approach of the \emph{ReStore}  system can be viewed as analogous to the steps of building and using materialized views for relational databases~\cite{ViewSurvey}.

\subsection{Support of Data Indices and Column Storage}
One of the main limitations of the original implementation of the MapReduce framework is that it is designed in a way that the jobs can only scan the input data in a sequential-oriented fashion. Hence, the query processing performance of the MapReduce framework  is unable to match the performance of a well-configured parallel DBMS~\cite{HadoopDB2}. In order to tackle this challenge,~\cite{HadoopPlus} have presented the \emph{Hadoop++ }system which aims to boost the  query performance of the Hadoop system without changing any of the system internals. They achieved this goal by injecting their changes  through user-defined function (UDFs) which only affect the Hadoop system from inside without any external effect. In particular, they introduce the following main changes:
\begin{compactitem}
\item \emph{Trojan Index}:  The original Hadoop implementation does not provide index access due to the lack of a priori knowledge of the schema and the MapReduce jobs being executed. Hence, the Hadoop++ system is based on the assumption that if we know the schema and the anticipated MapReduce jobs, then we can create appropriate indices for the Hadoop tasks. In particular, trojan index is an approach to integrate indexing capability into Hadoop in a non-invasive way. These indices are created during the data loading time and thus have no penalty at query time.  Each trojan Index provides an optional index access path which can be used for selective MapReduce jobs. The scan access path can still be used for other MapReduce jobs. These indices are created by injecting appropriate UDFs inside the Hadoop implementation. Specifically, the main features of trojan indices can be summarized as follows:
    \begin{compactitem}
    \item \emph{No External Library or Engine}: Trojan indices integrate indexing capability natively into the Hadoop framework without imposing a distributed SQL-query engine on top of it.
    \item \emph{Non-Invasive}: They do not change the existing Hadoop framework. The index structure is implemented by providing the right UDFs.
    \item \emph{Optional Access Path}: They provide an optional index access path which can be used for selective MapReduce jobs. However, the scan access path can still be used for other MapReduce jobs.
    \item \emph{Seamless Splitting}: Data indexing adds an index overhead for each data split. Therefore, the logical split includes the data as well as the index as it automatically splits the indexed data at logical split boundaries.
    \item \emph{Partial Index}: Trojan Index need not be built on the entire split. However, it can be built on any contiguous subset of the split as well.
    \item \emph{Multiple Indexes}: Several Trojan Indexes can be built on the same split. However, only one of them can be the primary index. During query processing, an appropriate index can be chosen for data access based on the logical query plan and the cost model.
\end{compactitem}

\item \emph{Trojan Join}:  Similar to the idea of the trojan index, the Hadoop++ system assumes that if we know the schema and the expected workload, then we can co-partition the input data during the loading time. In particular, given any two input relations, they apply the same partitioning function on the join attributes of both the relations at data loading time and place the co-group pairs, having the same join key from the two relations, on the same split and hence on the same node. As a result, join operations can be then processed locally within each node at query time. Implementing the trojan joins do not require any changes to be made to  the existing implementation of the Hadoop framework. The only changes are made on the internal management of the data splitting process. In addition, trojan indices can be freely combined with trojan joins.
\end{compactitem}

The design and implementation of a column-oriented and binary backend storage format for Hadoop has been presented in~\cite{COLMR}.
In general, a straightforward way to implement a column-oriented storage format for Hadoop is to store each column of the input dataset in a separate file. However, this raises two main challenges:
\begin{compactitem}
\item It requires generating roughly equal sized splits so that a job can be effectively parallelized over the cluster.
\item It needs to ensure that the corresponding values from different columns in the dataset are co-located on the same node running the map task.
\end{compactitem}

The first challenge can be tackled by horizontally partitioning the dataset and storing each partition in a separate subdirectory.
The second challenge is harder to tackle because of the default 3-way block-level replication strategy of HDFS that provides fault tolerance on commodity servers but does not provide any co-location guarantees. ~\cite{COLMR} tackle this challenge by implementing a modified HDFS block placement policy which guarantees that the files corresponding to the different columns of a split are always co-located across replicas. Hence, when reading a dataset, the column input format can actually assign one or more split-directories to a single split and the column files of a split-directory are scanned sequentially where the records are reassembled using values from corresponding positions in the files.  A lazy record construction technique is used to mitigate the deserialization overhead in Hadoop, as well as eliminate unnecessary disk I/O. The basic idea behind lazy record construction is to deserialize only those columns of a record that are actually accessed in a map function. Each column of the input dataset can be compressed using one of the following compression schemes:

\begin{compactenum}
\item \emph{Compressed Blocks}: This scheme uses a standard compression algorithm to compress a block of contiguous column values. Multiple compressed blocks may fit into a single HDFS block. A header indicates the number of records in a compressed block and the block's size. This allows the block to be skipped if no values are accessed in it. However, when a value in the block is accessed, the entire block needs to be decompressed.

\item \emph{Dictionary Compressed Skip List}: This scheme is tailored for map-typed columns. It takes advantage of the
fact that the keys used in maps are often strings that are drawn from a limited universe. Such strings are well suited for dictionary compression. A dictionary is built of keys for each block of map values and store the compressed keys in a map using a skip list format. The main advantage of this scheme is that a value can be accessed without having to decompress an entire block of values.
\end{compactenum}

One advantage of this approach is that adding a column to a dataset is not an expensive operation. This can be done by simply placing an additional file for the new column in each of the split-directories. On the other hand, a potential disadvantage of this approach is that the available parallelism may be limited for smaller datasets. Maximum parallelism is achieved for a MapReduce job when
the number of splits is at least equal to the number of map tasks.

\begin{figure}[t]
  \centering
  \includegraphics[width=0.48\textwidth]{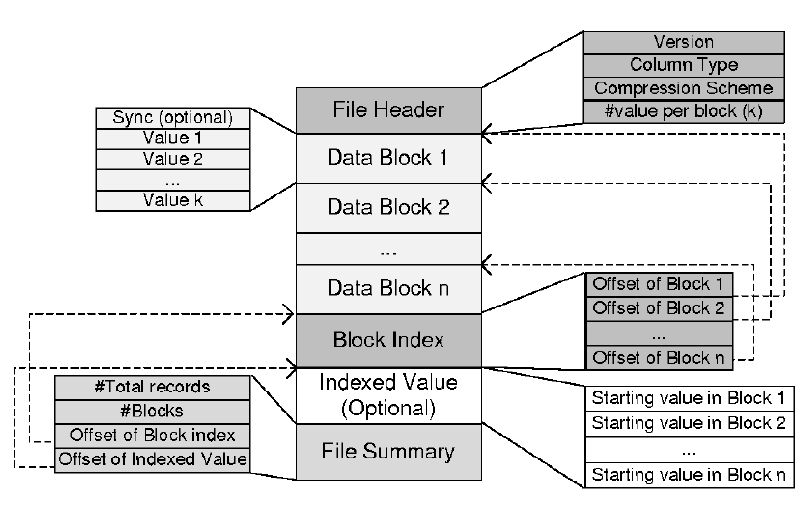}\\
  \caption{An Example Structure of \emph{CFile}~\cite{Llama}.}\label{Fig:Llama}
\end{figure}

The \emph{Llama} system~\cite{Llama} have introduced another approach of providing column storage support for the MapReduce framework. In this approach, each imported table is transformed into column groups where each group contains a set of files representing one or more columns. Llama introduced a column-wise format for Hadoop, called \emph{CFile}, where each file can contain multiple data blocks and each block of the file contains a fixed number of records (Figure~\ref{Fig:Llama}). However, the size of each logical block may vary since records can be variable-sized. Each file includes a block index, which is stored after all data blocks,  stores the offset of each block and is used to locate a specific block. In order to achieve storage efficiency, Llama uses block-level compression by using any of the well-known compression schemes. In order to improve the query processing and the performance of join operations, Llama columns are formed into correlation groups to provide the basis for the vertical partitioning of tables. In particular, it creates multiple vertical groups where each group is defined by a collection of columns, one of them is specified as the sorting column. Initially, when a new table is imported into the system, a basic vertical group is created which contains all the columns of the table and sorted by the table's primary key by default. In addition, based on statistics of query patterns, some auxiliary groups are dynamically created or discarded to improve the query performance.
The \emph{Clydesdale} system~\cite{Clydesdale,Clydesdale2}, a system which has been implemented for targeting workloads where the data fits a star schema, uses \emph{CFile} for storing its fact tables. It also relies on tailored join plans
and block iteration mechanism~\cite{BlockIteration} for optimizing the execution of its target workloads.

\emph{RCFile}~\cite{RCFile} (Record Columnar File) is another data placement structure that provides column-wise storage for Hadoop file system (HDFS). In RCFile, each table is firstly stored as horizontally partitioned into multiple row groups where each row group is then vertically partitioned so that each column is stored independently. 
In particular, each table can have multiple HDFS blocks where each block organizes records with the basic unit of a row group.
Depending on the row group size and the HDFS block size, an HDFS block can have only one or multiple row groups. In particular, a row group contains the following three sections:
\begin{compactenum}
\item The \emph{sync marker} which is placed in the beginning of the row group and mainly used to separate two continuous row groups in an HDFS block.
\item A metadata header which stores the information items on how many records are in this row group, how many bytes are in each column and how many bytes are in each field in a column.
\item The table data section which is actually a column-store where all the fields in the same column are stored continuously together.
\end{compactenum}
RCFile utilizes a column-wise data compression within each row group and provides a lazy decompression technique to avoid unnecessary column decompression during query execution.
In particular, the metadata header section is compressed using the \emph{RLE} (Run Length Encoding) algorithm.
The table data section is not compressed as a whole unit. However, each column is independently compressed with the \emph{Gzip} compression algorithm.
When processing a row group, RCFile does not need to fully read the whole content of the row group into memory.
However, it only reads the metadata header and the needed columns in the row group for a given query and thus it can skip unnecessary columns and gain the I/O advantages of a column-store.
The metadata header is always decompressed and held in memory until RCFile processes the next row group. However, RCFile does
not decompress all the loaded columns and uses a lazy decompression technique where a column will not be decompressed in memory until RCFile has determined that the data in the column will be really useful for query execution.

The notion of \emph{Trojan Data Layout} has been coined in~\cite{TrojanLayout} which exploits the existing data block replication in
HDFS to create different Trojan Layouts on a per-replica basis. This means that rather than keeping all data block replicas in the same layout, it uses \emph{different} Trojan Layouts for each replica which is optimized for a different subclass of queries.
As a result, every incoming query can be scheduled to the most suitable data block replica.
In particular, Trojan Layouts change the internal organization of a data block and not among
data blocks. They co-locate attributes together according to query workloads by applying a column grouping algorithm which
uses an interestingness measure that denotes how well a set of attributes speeds up most or all queries in a workload.
The column groups are then packed in order to maximize the total interestingness of data blocks.
At query time, an incoming MapReduce job is transparently adapted to query the data block replica that minimizes the data access
time. The map tasks are then routed of the MapReduce job to the data nodes storing such data block replicas.

\subsection{Effective Data Placement}
\begin{figure}[t]
  \centering
  \includegraphics[width=0.5\textwidth]{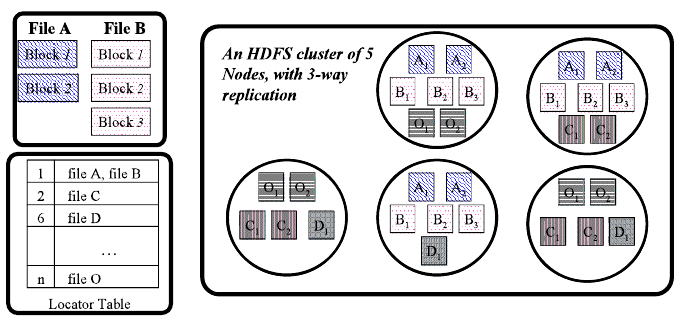}\\
  \caption{Example file colocation in CoHadoop~\cite{CoHadoop}. }\label{Fig:CoHadoop}
\end{figure}

In the basic implementation of the Hadoop project, the objective of the data placement policy is to achieve good load balance by distributing the data evenly across the data servers, independently of the intended use of the data. This simple data placement policy works well with most Hadoop applications that access just a \emph{single} file. However, there are some other applications that process data from \emph{multiple} files which can get a significant boost in performance with customized strategies. In these applications, the absence of data colocation increases the data shuffling costs, increases the network overhead and reduces the effectiveness of data partitioning. 
\emph{CoHadoop}~\cite{CoHadoop} is a lightweight extension to Hadoop which is designed to enable colocating related files at the file system level while at the same time retaining the good load balancing and fault tolerance properties. It introduces a new file property to identify related data files and modify the data placement policy of Hadoop to colocate copies of those related files in the same server. These changes are designed in a way to retain the benefits of Hadoop, including load balancing and fault tolerance. In principle,  CoHadoop provides a generic mechanism that allows applications to control data placement at the file-system level.  In particular, a new file-level property called a \emph{locator} is introduced and the Hadoop's data placement policy is modified so that it makes use of this locator property. Each locator is represented by a unique value (ID) where each file in HDFS is assigned to at most one locator and many files can be assigned to the same locator. Files with the same locator are placed on the same set of datanodes, whereas files with no locator are placed via Hadoop's default strategy. It should be noted that this colocation process involves all data blocks, including replicas. Figure \ref{Fig:CoHadoop} shows an example of colocating two files, \emph{A} and \emph{B}, via a common locator. All of \emph{A}'s two HDFS blocks and \emph{B}'s three blocks are stored on the same set of datanodes. To manage the locator information and keep track of collocated files, CoHadoop introduces a new data structure, \emph{the locator table}, which stores a mapping of locators to the list of files that share this locator.  In practice, the CoHadoop extension enables a wide variety of applications to exploit data colocation by simply specifying related files such as: colocating log files with reference files for joins, collocating partitions for grouping and aggregation, colocating index files with their data files and colocating columns of a table.

\subsection{Pipelining and Streaming Operations}
The original implementation of the  MapReduce framework has been designed in a way that the entire output of each map and reduce task to be \emph{materialized} into a local file before it can be consumed by the next stage. This materialization step allows for the implementation of a simple and elegant checkpoint/restart fault tolerance mechanism. The \emph{MapReduce Online} approach.~\cite{MROnline,MROnline2} have been proposed as a modified  architecture of the MapReduce framework in which intermediate data is \emph{pipelined} between operators while preserving the programming interfaces and fault tolerance models of previous MapReduce frameworks. This pipelining approach provides important advantages to the
MapReduce framework such as:
\begin{compactitem}
\item The reducers can begin their processing of the data as soon as it is produced by mappers. Therefore, they can generate and refine an approximation of their final answer during the course of execution. In addition, they can provide initial estimates of the results several orders of magnitude faster than the final results.
\item It widens the domain of problems to which MapReduce can be applied. For example, it facilitates the ability to design MapReduce jobs that run continuously, accepting new data as it arrives and analyzing it immediately (continuous queries). This allows MapReduce to be used for applications such as event monitoring and stream processing.
\item Pipelining delivers data to downstream operators more promptly, which can increase opportunities for parallelism, improve utilization and reduce response time.
\end{compactitem}

In this approach, each reduce task contacts every map task upon initiation of the job and opens a TCP socket which will be used to pipeline the output of the map function. As each map output record is produced, the mapper determines which partition (reduce task) the record should be sent to, and immediately sends it via the appropriate socket. A reduce task accepts the pipelined data it receives from each map task and stores it in an in-memory buffer. Once the reduce task learns that every map task has completed, it performs a final merge of all the sorted runs.
In addition, the reduce tasks of one job can optionally pipeline their output directly to the map tasks of the next job, sidestepping the need for expensive fault-tolerant storage in HDFS for what amounts to a temporary file. However, the computation of the reduce function from the previous job and the map function of the next job cannot be overlapped as the final result of the reduce step cannot be produced until all map tasks have completed, which prevents effective pipelining.
Therefore,  the reducer treats the output of a pipelined map task as \emph{tentative} until the JobTracker informs the reducer that the map task has committed successfully. The reducer can merge together spill files generated by the same uncommitted mapper, but will not combine those spill files with the output of other map tasks until it has been notified that the map task has committed. Thus, if a map task fails, each reduce task can ignore any tentative spill files produced by the failed map attempt. The JobTracker will take care of scheduling a new map task attempt, as in standard Hadoop. In principle, the main limitation of the \emph{MapReduce Online} approach is that it is based on HDFS. Therefore, it is not suitable for streaming applications, in which data streams have to be processed without any disk involvement. A similar approach has been presented in~\cite{ADHOC} which defines an \emph{incremental} MapReduce job as
one that processes data in large batches of tuples and runs continuously according to a specific  window range and slide of increment. In particular, it produces a MapReduce result that includes all data within a window (of time or data size) every slide and considers landmark MapReduce jobs where the trailing edge of the window is fixed  and the system incorporates new data into the existing result.
Map functions are trivially continuous, and process data on a tuple-by-tuple basis. However, before the reduce function may process the mapped data, the data must be partitioned across the reduce operators and sorted. When the map operator first receives a new key-value pair, it calls the map function and inserts the result into the latest increment in the map results. The operator then assigns output key-value pairs to reduce tasks, grouping them according to the partition function.
Continuous reduce operators participate in the sort as well, grouping values by their keys before calling the reduce function.

The \emph{Incoop} system~\cite{Incoop} has been introduced as a MapReduce implementation  that has been adapted for incremental computations which detects the changes on the input datasets and enables the automatic update of the outputs of the MapReduce jobs by employing a fine-grained result reuse mechanism. In particular, it allows MapReduce programs which are not designed for incremental processing to be executed transparently in an incremental manner.   To achieve this goal, the design of Incoop introduces new techniques that are incorporated into the Hadoop MapReduce framework. For example, instead of relying on HDFS to store the input to MapReduce jobs, Incoop devises a file system called \emph{Inc-HDFS} (Incremental HDFS) that provides mechanisms to identify similarities in the input data of consecutive job runs. In particular, Inc-HDFS splits the input into chunks whose boundaries depend on the file contents so that small changes to input do not change all chunk boundaries. Therefore, this partitioning mechanism can maximize the opportunities for reusing results from previous computations, while preserving compatibility with HDFS by offering the same interface and semantics.
In addition, Incoop controls the granularity of tasks so that large tasks can be divided into smaller subtasks that can be re-used
even when the large tasks cannot. Therefore, it introduces a new \emph{Contraction phase} that leverages \emph{Combiner} functions to reduce the network traffic by anticipating a small part of the processing done by the Reducer
tasks and control their granularity.
Furthermore, Incoop improves the effectiveness of memoization by implementing an affinity-based scheduler that applies a work stealing algorithm to minimize the amount of data movement across machines.
This modified scheduler strikes a balance between exploiting the locality of previously computed results and executing tasks on any available machine to prevent straggling effects.
On the runtime, instances of incremental Map tasks take advantage of previously stored results by querying the memoization server.
If they find that the result has already been computed, they fetch the result from the location of their memoized output
and conclude.
Similarly, the results of a Reduce task are remembered by storing them persistently and locally where a mapping from a collision-resistant hash of the input to the location of the output is inserted in the memoization server.

The \emph{DEDUCE} system~\cite{DEDUCE} has been presented as a middleware that attempts
to combine real-time stream processing with the capabilities of a large scale data analysis framework like MapReduce. In particular, it extends the \emph{IBM's System S} stream processing engine and augments its capabilities with those of the MapReduce framework.
In this approach, the input data set to the MapReduce operator can be either pre-specified at compilation time or could be provided at runtime as a punctuated list of files or directories. Once the input data is available, the MapReduce operator spawns a MapReduce job and produces a list of punctuated list of files or directories, which point to the
output data. Therefore, a MapReduce operator can potentially spawn multiple MapReduce jobs over the application lifespan but such jobs are spawned only when the preceding job (if any) has completed its execution. Hence, multiple jobs can be cascaded together to create a data-flow of MapReduce operators where the output from the MapReduce operators can be read to provide updates to the stream processing operators.

\subsection{System Optimizations}

In general, running a single program in a MapReduce framework may require tuning a number of parameters by users or system administrators.  The settings of these parameters control various aspects of job behavior during execution such as memory allocation and usage, concurrency, I/O optimization, and network bandwidth usage. The submitter of a Hadoop job has the option to set these parameters either using a program-level interface or through XML configuration files. For any parameter whose value
is not specified explicitly during job submission, default values, either shipped along with the system or specified by the system administrator, are used~\cite{AutomaticOptimization}. Users can run into performance problems because they do not know how to set these parameters correctly, or because they do not even know that these parameters exist.~\cite{WhatIF} have focused on the optimization opportunities presented by the large space of configuration parameters for these programs. They introduced a \emph{Profiler} component to collect detailed statistical information from unmodified MapReduce programs and a \emph{What-if }Engine for fine-grained cost estimation. In particular, the Profiler component is responsible for two main aspects:
\begin{compactenum}
\item Capturing information at the fine granularity of phases within the map and reduce tasks of a MapReduce job execution. This information is crucial to the accuracy of decisions made by the What-if Engine and the Cost-based Optimizer components.
\item Using dynamic instrumentation to collect run-time monitoring information from unmodified MapReduce programs. The dynamic nature means that monitoring can be turned on or off on demand.
\end{compactenum}
The What-if Engine's accuracy come from how it uses a mix of simulation and model-based estimation at the phase level of MapReduce job execution~\cite{HadoopPM,Starfish2,Starfish}. For a given MapReduce program, the role of the cost-based optimizer component is to enumerate and search efficiently through the high dimensional space of configuration parameter settings, making appropriate
calls to the What-if Engine, in order to find a good configuration setting, it clusters parameters into lower-dimensional subspaces such that the globally-optimal parameter setting in the high-dimensional space can be generated by composing the optimal settings found for the subspaces.
\emph{Stubby}~\cite{Stubby} has been presented as  a cost-based optimizer for MapReduce workflows that searches through the subspace of the full plan space that can be enumerated correctly and costed based on the information available in any given setting.
Stubby enumerates the plan space based on plan-to-plan transformations and an efficient search algorithm.

The \emph{Manimal} system~\cite{Manimal,Manimal2} is designed as a static analysis-style mechanism for detecting opportunities for applying relational style optimizations in MapReduce programs. Like most programming-language optimizers, it is a best-effort system where it does not guarantee that it will find every possible optimization and it only indicates an optimization when it is entirely safe to do so. In particular, the analyzer component of the system is responsible for examining the MapReduce program and sends the resulting optimization descriptor to the optimizer component.
In addition, the analyzer also emits an index generation program that can yield a B+Tree of the  input file.
The optimizer uses the optimization descriptor, plus a catalog of pre-computed indexes, to choose an optimized execution plan, called an execution descriptor. This descriptor, plus a potentially-modified copy of the user's original program, is then sent for execution on the Hadoop cluster.  These steps are performed transparently from the user where the submitted program does not need to be modified by the programmer in any way.
In particular, the main task of the analyzer is to produce a set of optimization descriptors which enable the system to carry out a phase roughly akin to logical rewriting of query plans in a relational database. The descriptors characterize a set of potential modifications that remain logically identical to the original plan.
The catalog is a simple mapping from a filename to zero or more ($X,O$) pairs where $X$ is an index file and $O$ is an optimization descriptor. The optimizer examines the catalog to see if there is any entry for input file. If not, then it simply indicates that Manimal should run the unchanged user program without any optimization. If there is at least one entry for the input file, and a catalog-associated optimization descriptor is compatible with analyzer-output, then the optimizer can choose an execution plan that takes advantage of the associated index file.

A key feature of MapReduce is that it automatically handles failures, hiding the complexity of fault-tolerance from the programmer. In particular, if a node crashes, MapReduce automatically restart the execution of its tasks. In addition, if a node is available but is performing poorly, MapReduce runs a speculative copy of its task (backup task) on another machine to finish the computation faster.
Without this mechanism of speculative execution, a job would be as slow as the misbehaving task. This situation can arise for many reasons, including faulty hardware and system misconfiguration.
On the other hand, launching too many speculative tasks may take away resources from useful tasks.
Therefore, the accuracy in estimating the progress and time-remaining long running jobs is an important challenge for a runtime environment like the MapReduce framework. In particular,  this information can play an important role in improving resource allocation, enhancing the task scheduling, enabling query debugging or tuning the cluster configuration. The \emph{ParaTimer} system~\cite{ParaTimer1,ParaTimer2} has been proposed to tackle this challenge. In particular,  ParaTimer provides techniques for handling several challenges including failures and data skew.
To handle unexpected changes in query execution times such as those due to failures, ParaTimer provides users with a set of time-remaining estimates that correspond to the predicted query execution times in different scenarios (i.e., a single worst-case failure, or data skew at an operator). Each of these indicators can be annotated with the scenario to which it corresponds, giving users a detailed picture of possible expected behaviors. To achieve this goal,
ParaTimer estimates time-remaining by breaking queries into pipelines where the time-remaining for each pipeline is estimated by considering the work to be done and the speed at which that work will be performed, taking (time-varying) parallelism into account. To get processing speeds, ParaTimer relies on earlier debug runs of the same query on input data samples generated by the user.
In addition, ParaTimer identifies the critical path in a query plan where it then estimates progress along that path, effectively ignoring other paths.~\cite{HeterogeneousMR} have presented an approach to estimate the progress of MapReduce tasks within environments of clusters with heterogenous hardware configuration. In these environments, choosing the node on which to run a speculative task is as important as choosing the task. They proposed an algorithm for speculative execution called \emph{LATE} (Longest Approximate Time to End) which is based on three principles: prioritizing tasks to speculate, selecting fast nodes on which to run and capping speculative tasks to prevent thrashing. In particular, the algorithm speculatively execute the task that it suspects will finish farthest into the future, because this task provides the greatest opportunity for a speculative copy to overtake the original and reduce the job's response time.  To really get the best chance of beating the original task with the speculative task, the algorithm only launches
speculative tasks on fast nodes (and not the first available node). The \emph{RAFT} (\emph{R}ecovery
\emph{A}lgorithms for \emph{F}ast-\emph{T}racking) system~\cite{RAFT2,RAFT} has been introduced, as a part of the \emph{Hadoop++ } system~\cite{HadoopPlus}, for tracking and recovering MapReduce jobs under task or node failures. In particular, RAFT uses two main checkpointing mechanisms: \emph{local checkpointing} and \emph{query metadata checkpointing}.
On the one hand, the main idea of local checkpointing is to utilize intermediate results, which are by default persisted by Hadoop, as checkpoints of ongoing task progress computation.  In general, map tasks spill buffered intermediate results to local disk whenever the output buffer is on the verge to overflow. RAFT exploits this spilling phase to piggy-back checkpointing metadata on the latest spill of each map task. For each checkpoint, RAFT stores a triplet of metadata that includes the  \emph{taskID} which represents a unique task identifier, \emph{spillID} which represents the local path to the spilled data and \emph{offset} which specifies the last byte of input data that was processed in that spill.
To recover from a task failure, the RAFT scheduler reallocates the failed task to the same node that was running the task. Then, the node resumes the task from the last checkpoint and reuses the spills previously produced for the same task. This simulates a situation where previous spills appear as if they were just produced by the task. In case that there is no local checkpoint available, the node recomputes the task from the beginning. On the other hand, the idea behind query metadata checkpointing is to push intermediate results to reducers as soon as map tasks are completed and to keep track of those incoming key-value pairs that produce local partitions and hence that are not shipped to another node for processing.
Therefore, in case of a node failure, the RAFT scheduler can recompute local partitions.

\section{Systems of Declarative Interfaces for the MapReduce Framework}
\label{SEC:SQLLike}
For programmers, a key appealing feature in the MapReduce framework is that there are only two main high-level declarative primitives (\emph{map} and \emph{reduce}) that can be written in any programming language of choice and without worrying about the details of their parallel execution. On the other hand, the MapReduce programming model has its own limitations such as:
\begin{compactitem}
\item Its one-input data format (key/value pairs) and two-stage data flow is extremely rigid. As we have previously discussed, to perform tasks that have a different data flow (e.g. joins or $n$ stages) would require the need to devise inelegant workarounds.
\item Custom code has to be written for even the most common operations (e.g. projection and filtering) which leads to the fact that the code is usually difficult to reuse and maintain unless the users build and maintain their own libraries with the common functions they use for processing their data.
\end{compactitem}

Moreover, many programmers could be unfamiliar with the MapReduce framework and they would prefer to use SQL (in which they are more proficient) as a high level declarative language to express their task while leaving all of the execution optimization details to the backend engine. In addition, it is beyond doubt that  high level language abstractions enable the underlying system to perform automatic optimization. In the following subsection we discuss research efforts that have been proposed to tackle these problems and add SQL-like interfaces on top of the MapReduce framework.

\subsection{Sawzall}

\begin{figure}[t]
    \begin{Verbatim}[frame=single]
    count: table sum of int;
    total: table sum of float;
    sumOfSquares: table sum of float;
    x: float = input;
    emit count $<$- 1;
    emit total $<$- x;
    emit sumOfSquares $<$- x * x;
  \end{Verbatim}
  \caption{An Example Sawzall Program}
\end{figure}

\emph{Sawzall}~\cite{Sawzall} is a scripting language used at Google on top of MapReduce. A Sawzall program defines the operations to be performed on a single record of the data. There is nothing in the language to enable examining multiple input records simultaneously, or even to have the contents of one input record influence the processing of another. The only output primitive in the language is the $emit$ statement, which sends data to an external aggregator (e.g. Sum, Average, Maximum, Minimum)  that gathers the results from each record after which the results are then correlated and processed. The authors argue that aggregation is done outside the language for a couple of reasons: 1) A more traditional language can use the language to correlate results but some of the aggregation algorithms are sophisticated and are best implemented in a native language and packaged in some form. 2) Drawing an explicit line between filtering and aggregation enables a high degree of parallelism and hides the parallelism from the language itself.

Figure \ref{Fig:Sawzall} depicts an example Sawzall program where the first three lines declare the aggregators \emph{count}, \emph{total} and \emph{sum of squares}. The keyword \emph{table} introduces an aggregator type which are called tables in Sawzall even though they may be singletons. These particular tables are \emph{sum} tables which add up the values emitted to them, \emph{ints} or \emph{floats} as appropriate. The Sawzall language is implemented as a conventional compiler, written in C++, whose target language is an interpreted instruction set, or byte-code. The compiler and the byte-code interpreter are part of the same binary, so the user presents source code to Sawzall and the system executes it directly. It is structured as a library with an external interface that accepts source code which is then compiled and executed, along with bindings to connect to externally-provided aggregators. The datasets of Sawzall programs are often stored in Google File System (GFS)~\cite{GFS}. The business of scheduling a job to run on a cluster of machines is handled by a software called \emph{Workqueue} which creates a large-scale time sharing system out of an array of computers and their disks. It schedules jobs, allocates resources, reports status and collects the results.

Google has also developed \emph{FlumeJava}~\cite{FlumeJava}, a Java library for developing and running data-parallel pipelines on top of MapReduce.  FlumeJava is centered around a few classes that represent parallel collections. Parallel collections support a modest number of parallel operations which are composed to implement data-parallel computations where an entire pipeline, or even multiple pipelines, can be translated into a single Java program using the FlumeJava abstractions.
To achieve good performance, FlumeJava internally implements parallel operations using \emph{deferred} evaluation. The invocation of a parallel operation does not actually run the operation, but instead simply records the operation and its arguments in an internal execution plan graph structure. Once the execution plan for the whole computation has been constructed, FlumeJava optimizes the execution
plan and then runs the optimized execution plan. When running the execution
plan, FlumeJava chooses which strategy to use to implement
each operation (e.g., local sequential loop vs. remote parallel MapReduce) based in part on the size of the data being processed,
places remote computations near the data on which they operate and performs independent operations in parallel.

\subsection{Pig Latin}
\begin{figure}[t]
  \centering
  \includegraphics[width=0.5\textwidth]{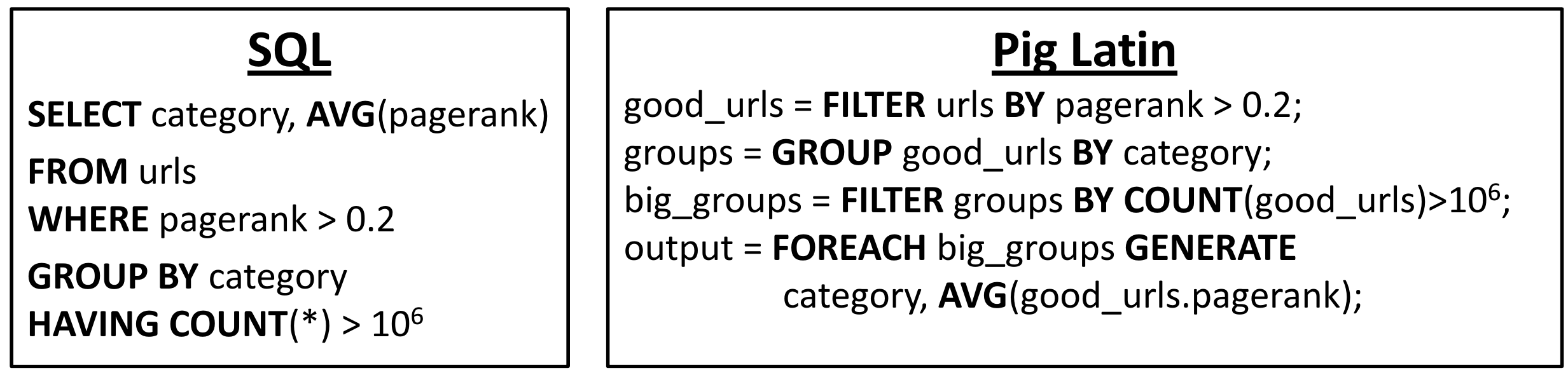}\\
  \caption{An Example SQL Query and Its Equivalent Pig Latin Program~\cite{PigLatin2}. }\label{Fig:PigLatin}
\end{figure}

Olston et al.~\cite{PigLatin1} have presented a language called \emph{Pig Latin} that takes a \emph{middle} position between expressing task using the high-level declarative querying model in the spirit of SQL and the low-level/procedural programming model using MapReduce. Pig Latin is implemented in the scope of the \emph{Apache Pig} project\footnote{http://incubator.apache.org/pig} and is used by programmers at Yahoo! for developing data analysis tasks.
Writing a Pig Latin program is similar to specifying a query execution plan (e.g. a data flow graph). To experienced programmers, this method is  more appealing than encoding their task as an SQL query and then coercing the system to choose the desired plan through optimizer hints. In general, automatic query optimization has its limits especially with uncataloged data, prevalent user-defined functions and parallel execution, which are all features of the data analysis tasks targeted by the MapReduce framework. Figure \ref{Fig:PigLatin} shows an example SQL query and its equivalent Pig Latin program. Given a $URL$ table with the structure $(url, category, pagerank)$, the task of the SQL query is to find each large category and its average pagerank of high-pagerank urls ($>$ 0.2). A Pig Latin program is described as a sequence of steps where each step represents a single data transformation. This characteristic is appealing to many programmers. At the same time, the transformation steps are described using high-level primitives (e.g. filtering, grouping, aggregation) much like in SQL.

Pig Latin has several other features that are important for casual ad-hoc data analysis tasks. These features include support for a flexible, fully nested data model, extensive support for user-defined functions and the ability to operate over plain input files without any schema information~\cite{ProgrammingPig}. In particular, Pig Latin has a simple data model consisting of the following four types:
\begin{compactenum}
\item \emph{Atom}: An atom contains a simple atomic value such as a string or a number, e.g. "alice".
\item \emph{Tuple}: A tuple is a sequence of fields, each of which can be any of the data types, e.g. ("alice", "lakers").
\item \emph{Bag}: A bag is a collection of tuples with possible duplicates. The schema of the constituent tuples is flexible where not all tuples in a bag need to have the same number and type of fields

    e.g.    $\left\{
    \begin{tabular}{l}
      ("alice", "lakers")\\
      ("alice", ("iPod", "apple"))\\
    \end{tabular}
     \right\} $
\item \emph{Map}: A map is a collection of data items, where each item has an associated key through which it can be looked up. As with bags, the schema of the constituent data items is flexible However, the keys are required to be data atoms, e.g.    $\left\{
    \begin{tabular}{l}
      "k1" $\rightarrow$ ("alice", "lakers")\\
      "k2" $\rightarrow$ "20"\\
    \end{tabular}
     \right\} $
\end{compactenum}

\begin{figure}[t]
  \centering
  \includegraphics[width=0.18\textwidth]{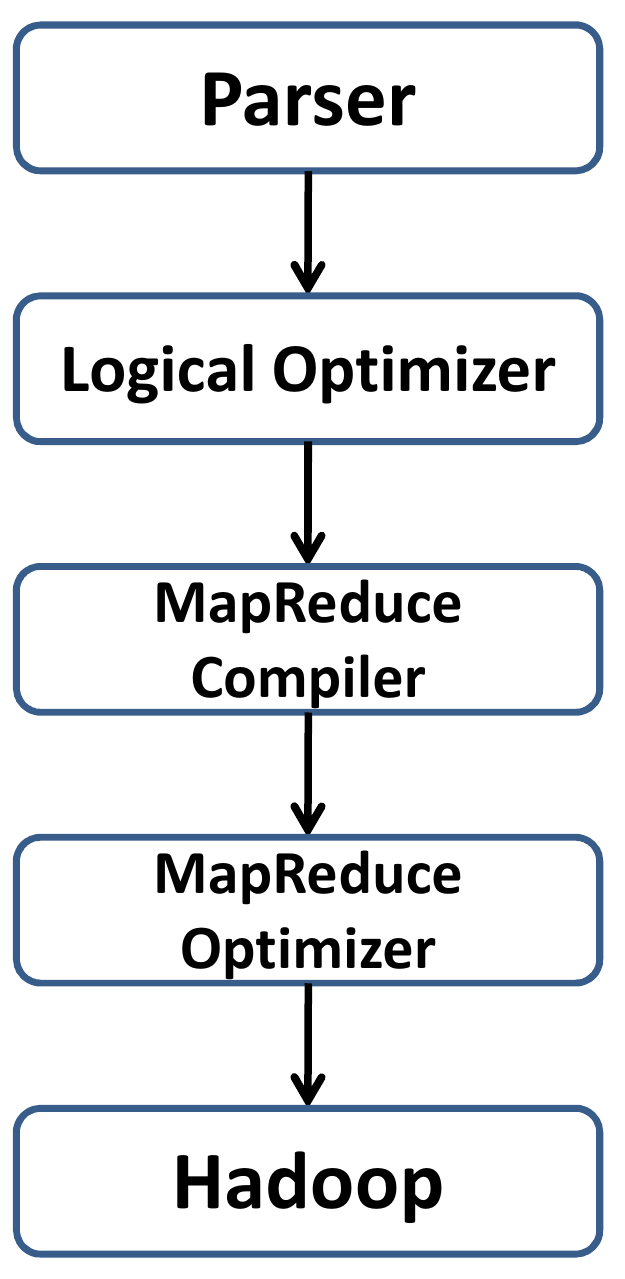}\\
  \caption{Pig compilation and execution steps~\cite{PigLatin1}. }\label{Fig:PigLatin2}
\end{figure}

\begin{figure*}[t]
\begin{Verbatim}[frame=single]
FROM (
    MAP doctext USING 'python wc_mapper.py' AS (word, cnt)
    FROM docs
    CLUSTER BY word
) a
REDUCE word, cnt USING 'python wc_reduce.py';
\end{Verbatim}
\caption{An Example HiveQl Query~\cite{Hive1}. }\label{Fig:HiveQl}
\end{figure*}
To accommodate specialized data processing tasks, Pig Latin has extensive support for user-defined functions (UDFs).  The input and output of UDFs in Pig Latin follow its fully nested data model. Pig Latin is architected such that the parsing of the Pig Latin program and the logical plan construction is independent of the execution platform. Only the compilation of the logical plan into a physical plan depends on the specific execution platform chosen. Currently, Pig Latin programs are compiled into sequences of MapReduce jobs which are executed using the Hadoop MapReduce environment. In particular, a Pig Latin program goes through a series of transformation steps~\cite{PigLatin1} before being executed as depicted in Figure~\ref{Fig:PigLatin2}. The parsing steps verifies that the program is syntactically correct and that all referenced variables are defined. The output of the parser is a canonical logical plan with a one-to-one correspondence between Pig Latin statements and logical operators which are arranged in a directed acyclic graph (DAG). The logical plan generated by the parser is passed through a logical optimizer. In this stage, logical optimizations such as projection pushdown are carried out. The optimized logical plan is then compiled into a series of MapReduce jobs which are then passed through another optimization phase.  The DAG of optimized MapReduce jobs is then topologically sorted and jobs are submitted to Hadoop for execution.

\subsection{Hive}
\label{SEC:HIVE}

The \emph{Hive} project\footnote{http://hadoop.apache.org/hive/} is an open-source data warehousing solution which has been built by the Facebook Data Infrastructure Team on top of the Hadoop environment~\cite{Hive1}. The main goal of this project is to bring the familiar relational database concepts (e.g. tables, columns, partitions) and a subset of SQL to the unstructured world of Hadoop while still maintaining the extensibility and flexibility that Hadoop provides. Thus, it supports all the major primitive types (e.g. integers, floats, strings) as well as complex types (e.g. maps, lists, structs).
Hive supports queries expressed in an SQL-like declarative language, \emph{HiveQL}\footnote{http://wiki.apache.org/hadoop/Hive/LanguageManual}, and therefore can be easily understood by anyone who is familiar with SQL. These queries are compiled into MapReduce jobs that are executed using Hadoop. In addition, HiveQL enables users to plug in custom MapReduce scripts into queries~\cite{HiveFB}. For example, the canonical MapReduce word count example on a table of documents (Figure \ref{Fig:MR1}) can be expressed in HiveQL as depicted in Figure~\ref{Fig:HiveQl} where the \emph{MAP} clause indicates how the input columns ($doctext$) can be transformed using a user program ('python wc\textunderscore mapper.py') into output columns ($word$ and $cnt$). The \emph{REDUCE} clause specifies the user program to invoke ('python wc\textunderscore reduce.py') on the output columns of the subquery.

HiveQL supports Data Definition Language (DDL) statements which can be used to create, drop and alter tables in a database~\cite{Hive2}. It allows users to load data from external sources and insert query results into Hive tables via the load and insert Data Manipulation Language (DML) statements respectively. However, HiveQL currently does not support the update and deletion of rows in existing tables (in particular, INSERT INTO, UPDATE and DELETE statements) which allows the use of very simple mechanisms to deal with concurrent read and write operations without implementing complex locking protocols. The metastore component is the Hive's system catalog which stores metadata about the underlying table. This metadata is specified during table creation and reused every time the table is referenced in HiveQL. The metastore distinguishes Hive as a traditional warehousing solution  when compared with similar data processing systems that are built on top of MapReduce-like architectures like Pig Latin~\cite{PigLatin1}.

\subsection{Tenzing}
The \emph{Tenzing} system~\cite{Tenzing} has been presented by Google as an SQL query execution engine which is built on top of MapReduce and provides a comprehensive SQL92 implementation with some SQL99 extensions (e.g. ROLLUP() and CUBE() OLAP extensions). Tenzing also supports querying data in different formats such as: row stores (e.g. MySQL database), column stores, \emph{Bigtable} (Google's built in key-value store)~\cite{BigTable}, \emph{GFS} (Google File System)~\cite{GFS}, text and protocol buffers. In particular, the Tenzing system has four major components:

\begin{compactitem}
\item \emph{The distributed worker pool}: represents the execution system which takes a query execution plan and executes the MapReduce jobs. The pool consists of master and worker nodes plus an overall gatekeeper called the master watcher. The workers manipulate the data for all the tables defined in the metadata layer.

\item \emph{The query server}: serves as the gateway between the client and the pool. The query server parses the query, applies different optimization mechanisms and sends the plan to the master for execution. In principle, the Tenzing optimizer applies some basic rule and cost-based optimizations to create an optimal execution plan.

\item \emph{Client interfaces}: Tenzing has several client interfaces including a command line client (CLI) and a Web UI. The CLI is a more powerful interface that supports complex scripting while the Web UI supports easier-to-use features such as query and table browsers tools. There is also an API to directly execute queries on the pool and a standalone binary which does not need any server side components but rather can launch its own MapReduce jobs.

\item \emph{The metadata server}: provides an API to store and fetch metadata such as table names and schemas and pointers to the underlying data.
\end{compactitem}

A typical Tenzing query is submitted to the query server (through the Web UI, CLI or API)
which is responsible for parsing the query into an intermediate
parse tree and fetching the required metadata from the metadata server. The query optimizer goes through the intermediate format,
 applies various optimizations and generates a query execution plan that consists of one or more
MapReduce jobs. For each MapReduce, the query server finds an available master using the master watcher and submits the query to it. At this stage, the execution is physically partitioned into multiple units of work where idle workers poll the masters for available work.  The query server monitors the generated intermediate results, gathers them as they arrive and streams the output back to the client.
In order to increase throughput, decrease latency and execute SQL operators more efficiently, Tenzing has enhanced the MapReduce implementation with the following main changes:

\begin{compactitem}
\item \emph{Streaming and In-memory Chaining}: The implementation of Tenzing does not serialize the intermediate results of MapReduce jobs to GFS. Instead, it streams the intermediate results between the Map and Reduce tasks using the network and  uses GFS only for backup purposes. In addition, it uses memory chaining mechanism where the reducer and the mapper of the same intermediate results are co-located in the same process.

\item \emph{Sort Avoidance}: Certain operators such as hash join and hash aggregation require shuffling but not sorting. The MapReduce API was enhanced to automatically turn off sorting for these operations, when possible, so that the mapper feeds data to the reducer which automatically bypasses the intermediate sorting step.  Tenzing also implements a block-based shuffle mechanism that combines many small rows into compressed blocks which is treated as one row in order to avoid reducer side sorting and avoid some of the overhead associated with row serialization and deserialization in the underlying MapReduce framework code.
\end{compactitem}

\subsection{SQL/MapReduce}

In general, a user-defined function (UDF) is a powerful database feature that allows users to customize database functionality. Friedman et al.~\cite{SQLMR} introduced the SQL/MapReduce (SQL/MR) UDF framework which is designed to facilitate parallel computation of procedural functions across hundreds of servers working together as a single relational database. The framework is implemented as part of the \emph{Aster Data Systems}\footnote{http://www.asterdata.com/} nCluster shared-nothing relational database. The framework leverage ideas from the MapReduce programming paradigm to provide users with a straightforward API through which they can implement a UDF in the language of their choice. Moreover, it allows maximum flexibility as the output schema of the UDF is specified by the function itself at query plan-time. This means that a SQL/MR function is polymorphic as it can process arbitrary input because its behavior as well as output schema are dynamically determined by information available at query plan-time. This also increases reusability as the same SQL/MR function can be used on inputs with many different schemas or with different user-specified parameters. In particular,  SQL/MR allows the user to write custom-defined functions in any programming language and insert them into queries that leverage traditional SQL functionality. A SQL/MR function is defined in a manner that is similar to MapReduce's map and reduce functions.

\begin{figure}[t]
  \centering
  \includegraphics[width=0.35\textwidth]{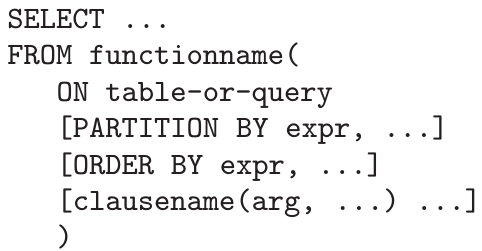}\\
  \caption{Basic Syntax of SQL/MR Query Function~\cite{SQLMR}.}\label{Fig:SQLMR}
\end{figure}

The syntax for using a SQL/MR function is depicted in Figure \ref{Fig:SQLMR} where the SQL/MR function invocation appears in the SQL \emph{FROM} clause and consists of the function name followed by a set of clauses that are enclosed in parentheses. The \emph{ON} clause specifies the input to the invocation of the SQL/MR function. It is important to note that the input schema to the SQL/MR function is specified implicitly at query plan-time in the form of the output schema for the query used in the ON clause.

In practice, a SQL/MR function can be either a mapper (\emph{Row} function) or a reducer (\emph{Partition} function). The definitions of row and partition functions ensure that they can be executed in parallel in a scalable manner. In the \emph{Row Function}, each row from the input table or query will be operated on by exactly one instance of the SQL/MR function. Semantically, each row is processed independently, allowing the execution engine to control parallelism. For each input row, the row function may emit zero or more rows. In the \emph{Partition Function}, each group of rows as defined by the \emph{PARTITION BY} clause will be operated on by exactly one instance of the SQL/MR function. If the \emph{ORDER BY} clause is provided, the rows within each partition are provided to the function instance in the specified sort order. Semantically, each partition is processed independently, allowing parallelization by the execution engine at the level of a partition. For each input partition, the SQL/MR partition function may output zero or more rows.

\subsection{HadoopDB}

There has been a long debate on the comparison between MapReduce framework and parallel database systems\footnote{http://databasecolumn.vertica.com/database-innovation/mapreduce-a-major-step-backwards/}~\cite{MapReduce4}. Pavlo et al.~\cite{HadoopDB2} have conducted a large scale comparison between the Hadoop implementation of MapReduce framework and parallel SQL database management systems in terms of performance and development complexity. The results of this comparison have shown that parallel database systems displayed a significant performance advantage over  MapReduce in executing a variety of data intensive analysis tasks. On the other hand, the Hadoop implementation was very much easier and more straightforward to set up and use in comparison to that of the parallel database systems. MapReduce have also shown to have superior performance in minimizing the amount of work that is lost when a hardware failure occurs. In addition, MapReduce (with its open source implementations) represents a very cheap solution in comparison to the very financially expensive parallel DBMS solutions (the price of an installation of a parallel DBMS cluster usually consists of 7 figures of U.S. Dollars)\cite{MapReduce4}.

\begin{figure}[t]
  \centering
  \includegraphics[width=0.5\textwidth]{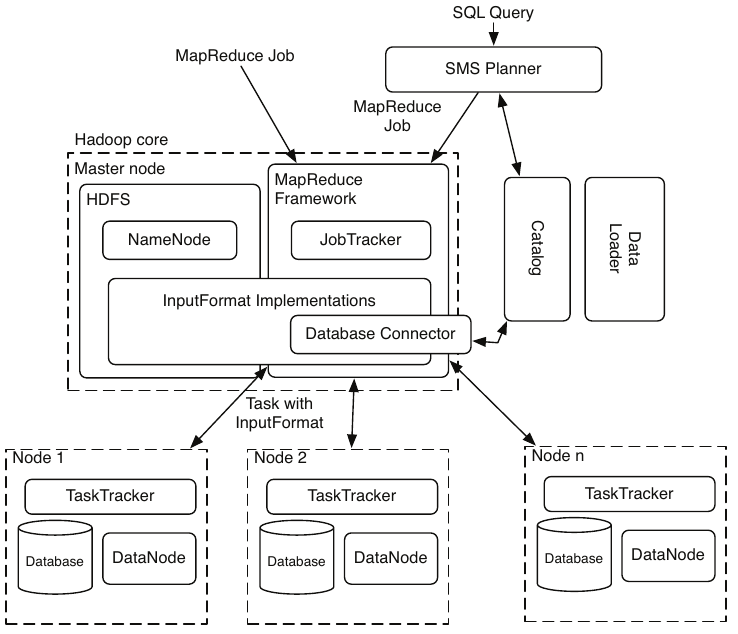}\\
  \caption{The Architecture of HadoopDB~\cite{HadoopDB}.}\label{Fig:HadoopDB}
\end{figure}

The \emph{HadoopDB} project\footnote{http://db.cs.yale.edu/hadoopdb/hadoopdb.html} is a hybrid system that tries to combine the scalability advantages of MapReduce with the performance and efficiency advantages of parallel databases~\cite{HadoopDB}. The basic idea behind HadoopDB is to connect multiple single node database systems (PostgreSQL) using Hadoop as the task coordinator and network communication layer. Queries are expressed in SQL but their execution are parallelized across nodes using the MapReduce framework, however, as much of the single node query work as possible is pushed inside of the corresponding node databases. Thus, HadoopDB tries to achieve fault tolerance and the ability to operate in heterogeneous environments by inheriting the scheduling and job tracking implementation from Hadoop. Parallely, it tries to achieve the performance of parallel databases by doing most of the query processing inside the database engine.
Figure \ref{Fig:HadoopDB} illustrates the architecture of HadoopDB which consists of two layers: 1) A data storage layer or the Hadoop Distributed File System\footnote{http://hadoop.apache.org/hdfs/} (HDFS). 2) A data processing layer or the MapReduce Framework. In this architecture, HDFS is a block-structured file system managed by a central \emph{NameNode}. Individual files are broken into blocks of a fixed size and distributed across multiple \emph{DataNodes} in the cluster. The NameNode maintains metadata about the size and location of blocks and their replicas. The MapReduce Framework follows a simple master-slave architecture. The master is a single \emph{JobTracker} and the slaves or worker nodes are \emph{TaskTrackers}. The JobTracker handles the runtime scheduling of MapReduce jobs and maintains information on each TaskTracker's load and available resources.  The \emph{Database Connector} is the interface between independent database systems residing on nodes in the cluster and TaskTrackers.  The Connector connects to the database, executes the SQL query and returns results as key-value pairs. The \emph{Catalog} component maintains metadata about the databases, their location, replica locations and data partitioning
properties. The \emph{Data Loader} component is responsible for globally repartitioning data on a given partition key upon loading and breaking apart single node data into multiple smaller partitions or chunks. The \emph{SMS planner} extends the HiveQL translator~\cite{Hive1} (Section \ref{SEC:HIVE}) and transforms SQL into MapReduce jobs that connect to tables stored as files in HDFS. Abouzeid et al.~\cite{HadoopDBDemo} have demonstrated HadoopDB in action running the following two different application types: 1) A semantic web application that provides biological data analysis of protein sequences. 2) A  classical business data warehouse.

\subsection{Jaql}
\label{SEC:Jaql}

Jaql\footnote{http://code.google.com/p/jaql/} is a query language which is designed for Javascript Object Notation (JSON)\footnote{http://www.json.org/}, a data format that has become popular because of its simplicity and modeling flexibility. JSON is a simple, yet flexible way to represent data that ranges from flat, relational data to semi-structured, XML data. Jaql is primarily used to analyze large-scale semi-structured data.  It is a functional, declarative query language which rewrites high-level queries when appropriate into a low-level query consisting of Map-Reduce jobs that are evaluated using the Apache Hadoop project. Core features include user extensibility and parallelism. Jaql consists of a scripting language and compiler, as
well as a runtime component~\cite{JAQL}. It is able to process data with no schema or only with a partial schema. However,
Jaql can also exploit rigid schema information when it is available, for both type checking and improved performance.
Jaql uses a very simple data model, a \emph{JDM value} is either an atom, an array or a record.
Most common atomic types are supported, including strings, numbers, nulls and dates.
Arrays and records are compound types that can be arbitrarily nested.
In more detail, an array is an ordered collection of values and can be used to model data structures such as vectors, lists, sets or bags.  A record is an unordered collection of name-value pairs and can model structs, dictionaries and maps.
Despite its simplicity, JDM is very flexible. It allows Jaql to operate with a variety of different data representations for both input and output, including delimited text files, JSON files, binary files, Hadoop's SequenceFiles, relational databases, key-value stores or XML documents.  Functions are first-class values in Jaql. They can be assigned to a variable and are high-order in that they can be passed as parameters or used as a return value. Functions are the key ingredient for reusability as any Jaql expression can be encapsulated in a function, and a function can be parameterized in powerful ways.

\begin{figure}[t]
  \centering
  \includegraphics[width=0.5\textwidth]{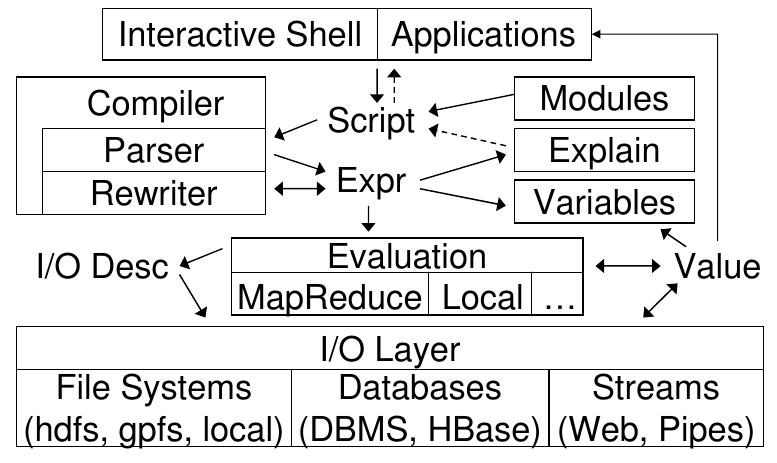}\\
  \caption{Jaql System Architecture~\cite{JAQL}.}\label{Fig:JAQL2}
\end{figure}

At a high-level, the Jaql architecture depicted in Figure \ref{Fig:JAQL2} is similar to most database systems.
Scripts are passed into the system from the interpreter or an application, compiled by the parser and rewrite engine, and either explained or evaluated over data from the I/O layer.
The storage layer is similar to a federated database. It provides an API to
access data of different systems including local or distributed
file systems (e.g., Hadoop's HDFS), database systems (e.g., DB2, Netezza, HBase), or from streamed sources like
the Web. Unlike federated databases, however, most of the accessed
data is stored within the same cluster and the I/O API describes
data partitioning, which enables parallelism with data affinity
during evaluation. Jaql derives much of this flexibility from
Hadoop's I/O API. It reads and writes many common file formats (e.g., delimited
files, JSON text, Hadoop Sequence files). Custom adapters are easily
written to map a data set to or from Jaql's data model. The input
can even simply be values constructed in the script itself. The Jaql interpreter evaluates the script locally on the computer
that compiled the script, but spawns interpreters on remote nodes
using MapReduce.  The Jaql compiler automatically detects parallelization opportunities in a Jaql script and translates it to a set of MapReduce jobs.

\section{Related Large Scale Data Processing Systems}
\label{SEC:SYSTEMS}
In this section, we give an overview of several large scale data processing systems that resemble some of the ideas of the MapReduce framework for different purposes and application scenarios. It must be noted, however, the design architectures and the implementations of these systems do not follow the architecture of the MapReduce framework and thus, they do not utilize and nor are they related to the infrastructure of the framework's open source implementations such as Hadoop.

\subsection{SCOPE}
\emph{SCOPE} (Structured Computations Optimized for Parallel Execution) is a scripting language which is targeted for large-scale data analysis and is used daily for a variety of data analysis and data mining applications inside Microsoft~\cite{SCOPE1}. SCOPE is a declarative language. It allows users to focus on the data transformations required to solve the problem at hand and hides the complexity of the underlying platform and implementation details. The SCOPE compiler and optimizer are responsible for generating an efficient execution plan and the runtime for executing the plan with minimal overhead.

\begin{figure}[t]
  \centering
  \includegraphics[width=0.5\textwidth]{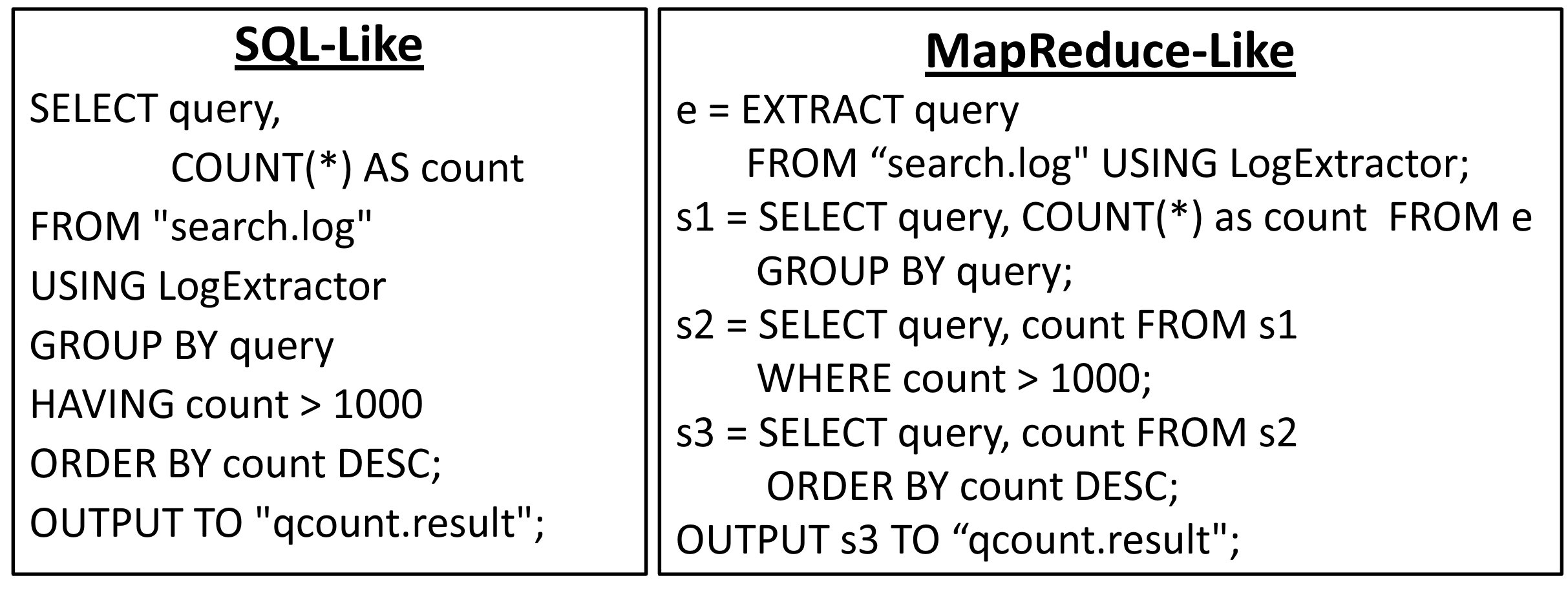}\\
  \caption{Two Equivalent SCOPE Scripts in SQL-like Style and MapReduce-Like Style~\cite{SCOPE1}.}\label{Fig:Scope2}
\end{figure}

Like SQL, data is modeled as sets of rows composed of typed columns.  SCOPE is highly extensible. Users can easily define their own functions and implement their own versions of operators: extractors (parsing and constructing rows from a file), processors (row-wise processing), reducers (group-wise processing) and combiners (combining rows from two inputs). This flexibility greatly extends the scope of the language and allows users to solve problems that cannot be easily expressed in traditional SQL. SCOPE provides a functionality which is similar to that of SQL views. This feature enhances modularity and code reusability. It is also used to restrict access to sensitive data. SCOPE supports writing a program using traditional SQL expressions or as a series of simple data transformations.  Figure~\ref{Fig:Scope2} illustrates two equivalent scripts in the two different styles (SQL-like and MapReduce-Like) to find from the search log the popular queries that have been requested at least 1000 times. In the MapReduce-Like style, the \emph{EXTRACT} command extracts all query string from the log file. The first \emph{SELECT} command counts the number of occurrences of each query string. The second \emph{SELECT} command retains only rows with a count greater than 1000. The third \emph{SELECT} command sorts the rows on count. Finally, the \emph{OUTPUT} command writes the result to the file "\emph{qcount.result}".

\begin{figure}[t]
  \centering
  \includegraphics[width=0.35\textwidth]{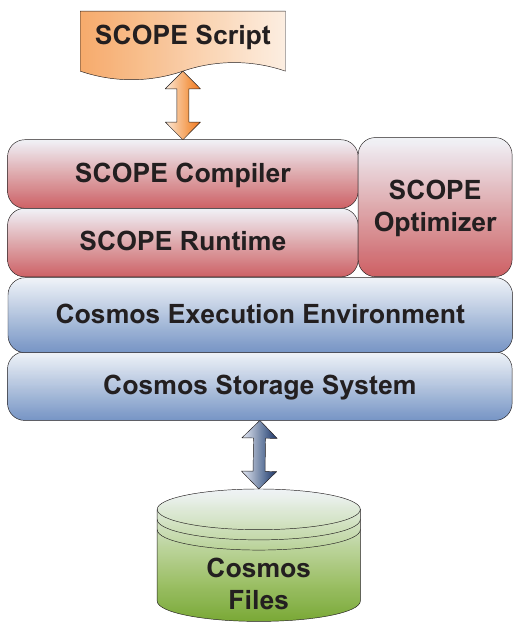}\\
  \caption{SCOPE/Cosmos Execution Platform~\cite{SCOPE1}.}\label{Fig:Scope}
\end{figure}

Microsoft has developed a distributed computing platform, called \emph{Cosmos}, for storing and analyzing massive data sets. Cosmos is designed to run on large clusters consisting of thousands of commodity servers. Figure \ref{Fig:Scope} shows the main components of the Cosmos platform which is described as follows:
\begin{compactitem}
\item  \emph{Cosmos Storage}: A distributed storage subsystem designed to reliably and efficiently store extremely large sequential files.
\item \emph{Cosmos Execution Environment}: An environment for deploying, executing and debugging distributed applications.
\item \emph{SCOPE}: A high-level scripting language for writing data analysis jobs. The SCOPE compiler and optimizer translate scripts to efficient parallel execution plans.
\end{compactitem}

The Cosmos Storage System is an append-only file system that reliably stores petabytes of data. The system is optimized for large sequential I/O. All writes are append-only and concurrent writers are serialized by the system. Data is distributed and replicated for fault tolerance and compressed to save storage and increase I/O throughput. In Cosmos, an application is modeled as a dataflow graph: a directed acyclic graph (DAG) with vertices representing processes and edges representing data flows. The runtime component of the execution engine is called the Job Manager which represents the central and coordinating process for all processing vertices within an application.

The SCOPE scripting language resembles SQL but with C$\#$ expressions. Thus, it reduces the learning curve for users and eases the porting of existing SQL scripts into SCOPE. Moreover, SCOPE expressions can use C$\#$ libraries where custom C$\#$ classes can compute functions of scalar values, or manipulate whole rowsets. A SCOPE script consists of a sequence of commands which are data transformation operators that take one or more rowsets as input, perform some operation on the data and output a rowset. Every rowset has a well-defined schema to which all its rows must adhere. The SCOPE compiler parses the script, checks the syntax and resolves names. The result of the compilation is an internal parse tree which is then translated to a physical execution plan. A physical execution plan is a specification of Cosmos job which describes a data flow DAG where each vertex is a program and each edge represents a data channel.  The translation into an execution plan is performed by traversing the parse tree in a bottom-up manner. For each operator, SCOPE has an associated default implementation rules.  Many of the traditional optimization rules from database systems are clearly also applicable  in this new context, for example, removing unnecessary columns, pushing down selection predicates and pre-aggregating when possible. However, the highly distributed execution environment offers new opportunities and challenges, making it necessary to explicitly consider the effects of large-scale parallelism during optimization. For example, choosing the right partition scheme and deciding when to partition are crucial for finding an optimal plan. It is also important to correctly reason about partitioning, grouping and sorting properties and their interaction, to avoid unnecessary computations~\cite{SCOPE2}.

\subsection{Dryad/DryadLinq} \emph{Dryad} is a general-purpose
distributed execution engine introduced by Microsoft for
coarse-grain data-parallel applications~\cite{Dryad1}. A Dryad
application combines computational \emph{vertices} with
communication \emph{channels} to form a dataflow graph. Dryad runs
the application by executing the vertices of this graph on a set
of available computers, communicating as appropriate through
files, TCP pipes and shared-memory FIFOs. The Dryad system allows
the developer fine control over the communication graph as well as
the subroutines that live at its vertices. A Dryad application
developer can specify an arbitrary directed acyclic graph to
describe the application's communication patterns and express the
data transport mechanisms (files, TCP pipes and shared-memory
FIFOs) between the computation vertices. This direct specification
of the graph gives the developer greater flexibility to easily
compose basic common operations, leading to a distributed analogue
of \emph{piping} together traditional Unix utilities such as grep,
sort and head.

Dryad is notable for allowing graph vertices to use an arbitrary number of inputs and outputs. The overall structure of a Dryad job is determined by its communication flow. A job is a directed acyclic graph where each vertex is a program and edges represent data channels. It is a logical computation graph that is automatically mapped onto physical resources by the runtime. At run time each channel is used to transport a finite sequence of structured items.  A Dryad job is coordinated by a process called the \emph{job manager} that runs either within the cluster or on a user's workstation with network access to the cluster. The job manager contains the application-specific code to construct the job's communication graph along with library code to schedule the work across the available resources. All data is sent directly between vertices and thus the job manager is only responsible for control decisions and is not a bottleneck for any data transfers. Therefore, much of the simplicity of the Dryad scheduler and fault-tolerance model come from the assumption that vertices are deterministic.

Dryad has its own high-level language called \emph{DryadLINQ}~\cite{Dryad2}. It generalizes execution environments such as SQL and MapReduce in two ways: 1) Adopting an expressive data model of strongly typed .NET objects. 2) Supporting general-purpose imperative and declarative operations on datasets within a traditional high-level programming language. DryadLINQ\footnote{http://research.microsoft.com/en-us/projects/dryadlinq/} exploits LINQ (Language INtegrated Query\footnote{http://msdn.microsoft.com/en-us/netframework/aa904594.aspx}, a set of .NET constructs for programming with datasets) to provide a powerful hybrid of declarative and imperative programming. The system is designed to provide flexible and efficient distributed computation in any LINQ-enabled programming language including C$\#$, VB and F$\#$\footnote{http://research.microsoft.com/en-us/um/cambridge/projects/fsharp/}. Objects in DryadLINQ datasets can be of any .NET type, making it easy to compute with data such as image patches, vectors and matrices. In practice, a DryadLINQ program is a sequential program composed of LINQ expressions that perform arbitrary side-effect-free transformations on datasets and can be written and debugged using standard .NET development tools. The DryadLINQ system automatically translates the data-parallel portions of the program into a distributed execution plan which is then passed to the Dryad execution platform.
A commercial implementation of Dryad and DryadLINQ was released in 2011 under the name \emph{LINQ to HPC}\footnote{http://msdn.microsoft.com/en-us/library/hh378101.aspx}.

\vspace{0.5cm}
\subsection{Spark}
The \emph{Spark} system~\cite{SPARK,SPARK2} have been proposed to support the applications which need to reuse
a working set of data across multiple parallel operations (e.g. iterative machine learning algorithms and interactive data analytic) while retaining the scalability and fault tolerance of MapReduce.
To achieve these goals, Spark introduces an abstraction called \emph{resilient distributed datasets} (RDDs).
An RDD is a read-only collection of objects partitioned across a set of machines that can be rebuilt if a partition is lost.
Therefore, users can explicitly cache an RDD in memory across machines and reuse it in multiple MapReduce-like parallel operations. RDDs do not need to be materialized at all times. RDDs achieve fault tolerance through a notion of \emph{lineage}. In particular,
each RDD object contains a pointer to its parent and information about how the parent was transformed.
Hence, if a partition of an RDD is lost, the RDD has sufficient information about how it was derived from other RDDs to be able to rebuild just that partition.

Spark is implemented in the Scala programming language\footnote{http://www.scala-lang.org/}~\cite{SCALA}. It is built on top of \emph{Mesos}~\cite{Mesos}, a cluster operating system that lets multiple parallel frameworks share a cluster in a fine-grained manner and provides an API for applications to launch tasks on a cluster.
It provides isolation and efficient resource sharing across frameworks running on the same cluster while giving each framework freedom to implement its own programming model and fully control the execution of its jobs.
Mesos uses two main abstractions: \emph{tasks} and \emph{slots}. A task represents a unit of work. A slot represents a computing resource in which a framework may run a task, such as a core and some associated memory on a multicore machine.
It employs the two-level scheduling mechanism. At the first level, Mesos allocates slots between frameworks using fair sharing.
At the second level, each framework is responsible for dividing its work into tasks, selecting which tasks to run in each slot. This lets frameworks perform application-specific optimizations. For example Spark's scheduler tries to send each task to one of its preferred locations using a technique called \emph{delay scheduling}~\cite{DelayScheduling}

To use Spark, developers need to write a driver program that implements the high-level control flow of their application and launches various operations in parallel.
Spark provides two main abstractions for parallel programming: resilient distributed datasets and parallel operations on these datasets (invoked by passing a function to apply on a dataset). In particular, each RDD is represented by a Scala object which can be constructed in different ways:
\begin{compactitem}
\item From a file in a shared file system (e.g HDFS).
\item By parallelizing a Scala collection (e.g., an array) in the driver program which means dividing it into a number of slices that will be sent to multiple nodes.
\item By transforming an existing RDD. A dataset with elements of type $A$ can be transformed into a dataset with
elements of type $B$ using an operation called $flatMap$.
\item By changing the persistence of an existing RDD. A user can alter the persistence of an RDD through two actions:
\begin{compactitem}
\item The cache action leaves the dataset lazy but hints that it should be kept in memory after the first time it is computed because it will be reused.
\item The save action evaluates the dataset and writes it to a distributed filesystem such as HDFS. The saved version is used in future operations on it.
\end{compactitem}
\end{compactitem}
Different parallel operations can be performed on RDDs:
\begin{compactitem}
\item The \emph{reduce} operation which combines dataset elements using an associative function to produce a result at the driver program.
\item The \emph{collect} operation which sends all elements of the dataset to the driver program.
\item The \emph{ foreach} which passes each element through a user provided function.
\end{compactitem}
Spark does not currently support a grouped reduce operation as in MapReduce. The results of reduce operations
are only collected at the driver process.

\subsection{Nephle/Pact}
The \emph{Nephele/PACT} system~\cite{Nephele,Nephele2} has been presented as a parallel data processor centered around a programming model of so-called \emph{Parallelization Contracts} (PACTs) and the scalable parallel execution engine \emph{Nephele}.
The PACT programming model is a generalization of map/reduce  as it is based on a key/value data model and the concept of \emph{Parallelization Contracts} (PACTs). A PACT consists of exactly one second-order function which is called \emph{Input Contract} and an optional \emph{Output Contract}. An Input Contract takes a first-order function with task-specific user code and one or more data sets as input parameters. The Input Contract invokes its associated first-order function with independent subsets of its input data in a data-parallel fashion. In this context, the two functions of \emph{map} and \emph{reduce} are just examples of the Input Contracts. Other example of Input Contracts include:
\begin{compactitem}
\item The \emph{Cross} contract which operates on multiple inputs and builds a distributed Cartesian product over its input
sets.
\item The \emph{CoGroup} contract partitions each of its multiple inputs along the key. Independent subsets are
built by combining equal keys of all inputs.
\item The \emph{Match} contract operates on multiple inputs. It matches key/value pairs from all input data sets with the same key (equivalent to the inner join operation).
\end{compactitem}

\begin{figure}[t]
  \centering
  \includegraphics[width=0.45\textwidth]{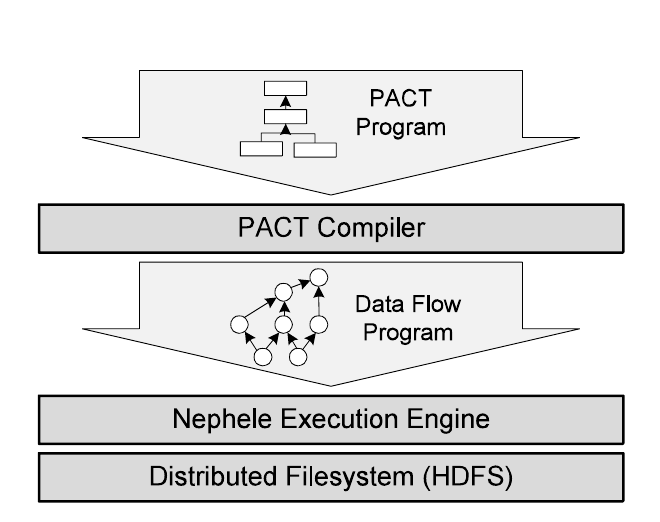}\\
  \caption{The Nephele/PACT System Architecture~\cite{Nephele2}.}\label{Fig:Nephele}
\end{figure}

An Output Contract is an optional component of a PACT and gives guarantees about the data that is generated by the assigned user function. The set of Output Contracts include:

\begin{compactitem}
\item The \emph{Same-Key} contract where each key/value pair that is generated by the function has the same key as the key/value pair(s) from which it was generated. This means the function will preserve any partitioning and order property on the keys.

\item The \emph{Super-Key} where each key/value pair that is generated by the function has a superkey of the key/value pair(s) from which it was generated. This means the function will preserve the partitioning and the partial order on the keys.
\item The \emph{Unique-Key} where each key/value pair that is produced has a unique key. The key must be unique across all parallel instances. Any produced data is therefore partitioned and grouped by the key.

\item The \emph{Partitioned-by-Key} where key/value pairs are partitioned by key. This contract has similar implications as the Super-Key contract, specifically that a partitioning by the keys is given, but there is no order inside the partitions.
\end{compactitem}

Figure~\ref{Fig:Nephele} illustrate the system architecture of Nephele/PACT where  a PACT program is submitted to the PACT Compiler which translates the program into a data flow execution plan which is then handed to the Nephele system for parallel execution.
Hadoop distributed filesystem (HDFS) is used for storing both the input and the output data.

Due to the declarative character of the PACT programming model, the PACT compiler can apply different optimization mechanisms and select from several execution plans with varying costs for a single PACT program.
For example, the \emph{Match} contract can be satisfied using either a repartition strategy which partitions all inputs by keys or a broadcast strategy that fully replicates one input to every partition of the other
input. Choosing the right strategy can dramatically reduce network traffic and execution time.
Therefore, the PACT compiler applies standard  SQL
optimization techniques~\cite{ACESSPATH} where it exploits information provided by the Output Contracts and apply different cost-based optimization techniques. In particular, the optimizer generates a set of candidate execution plans in a bottom-up fashion
(starting from the data sources) where  the more expensive plans are pruned using a set of \emph{interesting properties} for the operators. These properties are also used to spare plans from pruning that come with an additional property that may amortize their cost overhead later.

\subsection{Boom Analytics}
The  \emph{BOOM Analytics} (Berkeley Orders of Magnitude)~\cite{BoomAnalytics} is an API-compliant reimplementation of the HDFS
distributed file system (\emph{BOOM-FS}) and the Hadoop MapReduce engine (\emph{BOOM-MR}).
The implementation of BOOM Analytics uses the \emph{Overlog}
logic language~\cite{Overlog} which has been originally presented as an event-driven language
and evolved a semantics more carefully grounded in \emph{Datalog}, the standard deductive query language from database theory~\cite{DATALOG}. In general, the Datalog language is defined over relational tables as a purely logical query language that makes no changes to the stored tables. Overlog extends Datalog in three main features~\cite{Overlog2}:
\begin{compactenum}
\item It adds notation to specify the location of data.
\item It provides some SQL-style extensions such as primary keys and aggregation.
\item It defines a model for processing and generating changes to tables.
\end{compactenum}
When Overlog tuples arrive at a node either through rule evaluation or external events, they are handled in an atomic
local Datalog \emph{timestep}. Within a timestep, each node sees only locally-stored tuples. Communication between Datalog and the rest of the system (Java code, networks, and clocks) is modeled using events corresponding to insertions or deletions
of tuples in Datalog tables. BOOM Analytics uses a Java-based Overlog runtime called \emph{JOL} which compiles Overlog programs into pipelined dataflow graphs of operators. In particular, JOL provides metaprogramming support where each Overlog program is compiled into a representation that is captured in rows of tables.
In BOOM Analytics, \emph{everything} is data. This includes traditional persistent information like file system metadata, runtime state like TaskTracker status, summary statistics like those used by the JobTracker's scheduling policy, communication messages, system events and execution state of the system.

The BOOM-FS component represents the file system metadata as a collection of relations (\emph{file}, \emph{fqpath}, \emph{fchunk}, \emph{datanode}, \emph{hbchunk}) where file system operations are implemented by writing queries over these tables.
The \emph{file} relation contains a row for each file or directory stored in BOOM-FS.
The set of chunks in a file is identified by the corresponding rows in the \emph{fchunk} relation. The \emph{datanode}
and \emph{hbchunk} relations contain the set of live DataNodes and the chunks stored by each DataNode, respectively. The NameNode updates these relations as new heartbeats arrive. If the NameNode does not receive a heartbeat from a DataNode within a configurable amount of time, it assumes that the DataNode has failed and removes the corresponding rows from these tables.
Since a file system is naturally hierarchical, the file system queries that needed to traverse it are recursive.
Therefore, the parent-child relationship of files is used to compute the transitive closure of each file and store its fully-qualified path in the \emph{fqpath} relation.
Because path information is accessed frequently, the
  \emph{fqpath} relation is configured to be cached after it is computed.
Overlog will automatically update \emph{fqpath} when a file is changed, using standard relational view maintenance
logic~\cite{DATALOG}.  BOOM-FS also defines several views to compute derived file system metadata such as the total size of each
file and the contents of each directory. The materialization of
each view can be changed via simple Overlog table definition statements without altering the semantics of the program.
In general, HDFS  uses three different communication protocols: the \emph{metadata protocol}  which is used by clients and NameNodes to
exchange file metadata, the \emph{heartbeat protocol}  which is used by the DataNodes to notify the NameNode about chunk locations and DataNode liveness, and the \emph{data protocol} which is used by the  clients and
DataNodes to exchange chunks. BOOM-FS re-implemented these three protocols using a set of  Overlog rules.
BOOM-FS also achieves the high availability failover mechanism by using Overlog to implement the \emph{hot standby} NameNodes feature using  Lamport's Paxos algorithm~\cite{Lamport}

BOOM-MR re-implements the MapReduce framework by replacing Hadoop's core scheduling logic with Overlog.
The JobTracker tracks the ongoing status of the system and transient state in the form of messages sent and received by the JobTracker by capturing this information in four Overlog tables:  \emph{job}, \emph{task}, \emph{taskAttempt} and \emph{taskTracker}. The \emph{job} relation contains a single row for each job submitted to the JobTracker. The \emph{task} relation identifies each task within a job. The attributes of this relation identify the task type (map or reduce), the input partition (a chunk for map tasks, a bucket for reduce tasks) and the current running status.
The \emph{taskAttempt} relation maintains the state of each task attempt (A task may be attempted more than once due to speculation or if the initial execution attempt failed). The \emph{taskTracker} relation identifies each TaskTracker in the cluster with a unique name. Overlog rules are used to update the JobTracker's tables by converting inbound messages into tuples of the four Overlog tables. Scheduling decisions are encoded in the \emph{taskAttempt} table which assigns tasks to \emph{TaskTrackers}. A scheduling policy is simply a set of rules that join against the \emph{taskTracker} relation to find \emph{TaskTrackers} with unassigned slots and schedules tasks by inserting tuples into \emph{taskAttempt}. This architecture allows new scheduling policies to be defined easily.

\subsection{Hyracks/ASTERIX}
\begin{figure}[t]
  \centering
  \includegraphics[width=0.5\textwidth]{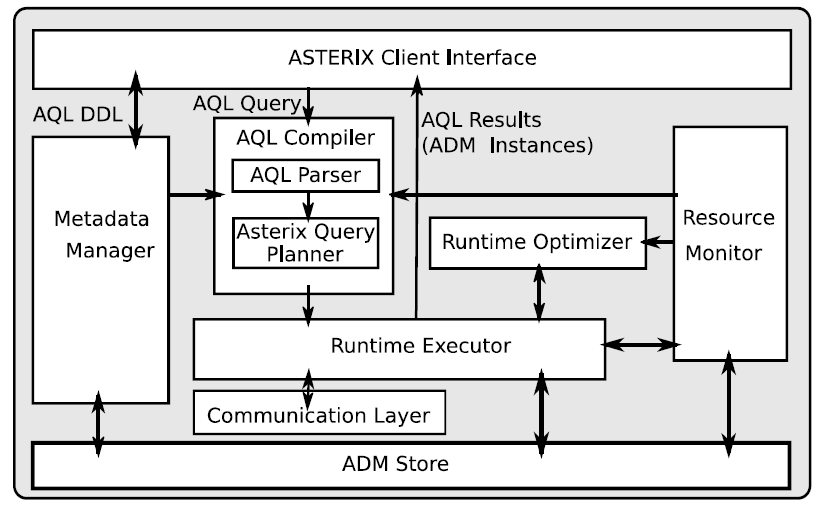}\\
  \caption{The ASTERIX System Architecture~\cite{ASTERIX}.}\label{Fig:ASTERIX}
\end{figure}
\emph{Hyracks} is presented as a partitioned-parallel dataflow execution platform that runs on shared-nothing clusters of computers~\cite{Hyracks}. Large collections of data items are stored as local partitions distributed across the nodes of the cluster. A Hyracks job is submitted by a client and processes one or more collections of data to produce one or more output collections (partitions).  Hyracks provides a programming model and an accompanying infrastructure to efficiently divide computations on large data collections (spanning multiple machines) into computations that work on
each partition of the data separately.  Every Hyracks cluster is managed by a \emph{Cluster Controller} process.  The Cluster Controller accepts job execution requests from clients, plans their evaluation strategies and then schedules the jobs' tasks to run on selected machines in the cluster.
In addition, it is responsible for monitoring the state of the cluster to keep track of the resource loads at the various worker machines.  The Cluster Controller is also responsible for re-planning and re-executing some or all of the tasks of a job in the event of a failure. On the task execution side, each worker machine that participates in a Hyracks cluster runs a \emph{Node Controller} process. The Node Controller accepts task execution requests from the Cluster Controller and also
reports on its health via a heartbeat mechanism.

In principle, Hyracks has been designed with the goal of being a runtime platform where users can create their jobs and also to serve as an efficient target for the compilers of higher-level programming languages such as Pig, Hive or Jaql.
The \emph{ASTERIX} project~\cite{ASTERIX,ASTERIX2} uses this feature with the aim of building a scalable information management system that supports the storage, querying and analysis of large collections of semi-structured nested data objects.
The ASTERIX data storage and query processing are based on its own semistructured model called the \emph{ASTERIX Data Model} (ADM). Each individual ADM data instance is typed and self-describing.
All data instances live in \emph{datasets} (the ASTERIX analogy to tables) and datasets can be indexed, partitioned and possibly
replicated to achieve the scalability and availability goals.
External datasets which reside in files that are not under ASTERIX control are also supported.
An instance of the ASTERIX data model can either be a primitive type (e.g., integer, string, time) or a derived type, which may
include:
\begin{compactitem}
\item \emph{Enum}: an enumeration type, whose domain is defined by listing the sequence of
possible values.
\item \emph{Record}: a set of fields where each field is described by its name and type. A record
can be either an open record where it contains fields that are not part of the type definition or a closed record which cannot.
\item \emph{Ordered list}: a sequence of values for which the order is determined by creation or
insertion.
\item \emph{Unordered list}: an unordered sequence of values which is similar to bags in SQL.
\item \emph{Union}: describes a choice between a finite set of types.
\end{compactitem}
A dataset is a target for AQL queries and updates and is also the attachment point for indexes.
A collection of datasets related to an application are grouped into a namespace called a \emph{dataverse} which is analogous to a database in the relational world.
In particular, data is accessed and manipulated through the use of the \emph{ASTERIX Query Language} (AQL) which is designed to cleanly match and handle the data structuring constructs of ADM.
It borrows from \emph{XQuery}  and \emph{Jaql} their programmer-friendly declarative syntax that describes bulk operations
such as iteration, filtering and sorting. Therefore, AQL is comparable to those languages
in terms of expressive power. The major difference with respect to XQuery is AQL's focus on data-centric use cases at the expense of built-in support for mixed content for document-centric use cases. In ASTERIX, there is no notion of document order or node identity for data instances. Differences between AQL and Jaql
stem from the usage of the languages. While ASTERIX data is stored in and managed by the ASTERIX system, Jaql runs against data that are stored externally in Hadoop files or in the local file system.
Figure~\ref{Fig:ASTERIX} presents an overview of the ASTERIX system architecture.
AQL requests are compiled into jobs for an ASTERIX execution layer, Hyracks.
ASTERIX concerns itself with the data details of AQL and ADM, turning AQL requests into Hyracks jobs while Hyracks determines and oversees the utilization of parallelism based on information and constraints associated with the resulting
jobs' operators as well as on the runtime state of the cluster.

\section{Conclusions}
\label{SEC:Conclusions}
The database community has been always focusing on dealing with the challenges of \emph{Big Data} management, although the meaning of "\emph{Big}" has been evolving continuously to represent different scales over the time~\cite{History}.
According to IBM, we are currently creating 2.5 quintillion bytes of data, everyday. This data comes from many different sources and in different formats including digital pictures, videos, posts to social media sites, intelligent sensors, purchase transaction records and cell phone GPS signals. This is a new scale of \emph{Big Data} which is attracting a huge interest from both the   industrial and research communities with the aim of creating the best means to process and analyze this data in order to make the best use of it.
In the last decade, the MapReduce framework has emerged as a popular mechanism to harness the power of large clusters of computers. It allows programmers to think in a \emph{data-centric} fashion where they can focus on applying transformations to sets of data records while the details of distributed execution and fault tolerance are transparently managed by the MapReduce framework.

In this article, we presented  a survey of the MapReduce family of approaches for developing scalable data processing systems and solutions.   In general we noticed that although the MapReduce framework, and its open source implementation of Hadoop, are now considered to be sufficiently mature such that they are widely used for developing many solutions by academia and industry in different application domains.  We believe that it is unlikely that MapReduce will substitute database systems even for data warehousing applications. We expect that they will always coexist and complement each others in different scenarios.
We are also convinced that there is still room for further optimization and advancement in different directions on the spectrum of the MapReduce framework that is required to bring forward the vision of providing large scale data analysis as a commodity for novice end-users. For example, energy efficiency in the MapReduce is an important problem which has not attracted enough attention from the research community, yet. The traditional challenge of debugging large scale computations on distributed system has not been  considered  as a research priority by the MapReduce research community. Related with the issue of the power of expressiveness of the programming model, we feel that this is an area that requires more investigation. We also noticed that the over simplicity of the MapReduce programming model have raised some key challenges on dealing with complex data models (e.g., nested models, XML and hierarchical model , RDF and graphs) efficiently. This limitation has called for the need of next-generation of big data architectures and systems that can provide the required scale and performance attributes for these domain. For example, Google has created the \emph{Dremel} system~\cite{Dremel}, commercialized under the name of \emph{BigQuery}\footnote{https://developers.google.com/bigquery/}, to support interactive analysis of nested data. Google has also presented the \emph{Pregel} system~\cite{Pregel}, open sourced by \emph{Apache Giraph} and \emph{Apache Hama} projects, that uses a BSP-based programming model for efficient and scalable processing of massive graphs on distributed clusters of commodity machines.
Recently, \emph{Twitter} has   announced the release of the \emph{Storm}\footnote{https://github.com/nathanmarz/storm/} system as a distributed and fault-tolerant platform for implementing continuous and realtime processing applications of streamed data. We believe that more of these domain-specific systems will be introduced in the future to form the new generation of big data systems. Defining the right and most convenient programming abstractions and declarative interfaces of these domain-specific Big Data systems is another important research direction that will need to be deeply investigated.

\bibliographystyle{plain}
\bibliography{Biblio}

\appendix
\section{Application of the MapReduce framework}
\label{SEC:CaseStudies}

MapReduce-based systems are increasingly being used for large-scale data analysis. There are several reasons for this such as~\cite{InDepthStudy}:
\begin{compactitem}
\item \emph{The interface of MapReduce is simple yet expressive.} Although MapReduce only involves two functions map and reduce, a number of data analytical tasks including traditional SQL query, data mining, machine learning and graph processing can be expressed with a set of MapReduce jobs.
\item \emph{MapReduce is flexible.} It is designed to be independent of storage systems and is able to analyze various kinds of data, structured and unstructured.
\item \emph{MapReduce is scalable}. Installation of MapReduce can run over thousands of nodes on a shared-nothing cluster while keeping to provide fine-grain fault tolerance whereby only tasks on failed nodes need to be restarted.
\end{compactitem}

These main advantages have triggered several research efforts with the aim of applying the MapReduce framework for solving challenging data processing problem on large scale datasets in different domains. In this appendix, we provide an overview of several research efforts of developing MapReduce-based solutions for data-intensive applications of different data models such as XML (Appendix~\ref{SEC:XMLMAPREDUCE}), RDF (Appendix~\ref{SEC:RDFMAPREDUCE})  and graphs (Appendix~\ref{SEC:GRAPHMAPREDUCE}). Appendix~\ref{SEC:OTHERMAPREDUCE} provides an overview of several approaches of developing MapReduce-based solutions in different data-intensive and computationally expensive operations.

\subsection{MapReduce for Large Scale XML Processing}
\label{SEC:XMLMAPREDUCE}
\begin{figure}[t]
  \centering
  \includegraphics[width=0.5\textwidth]{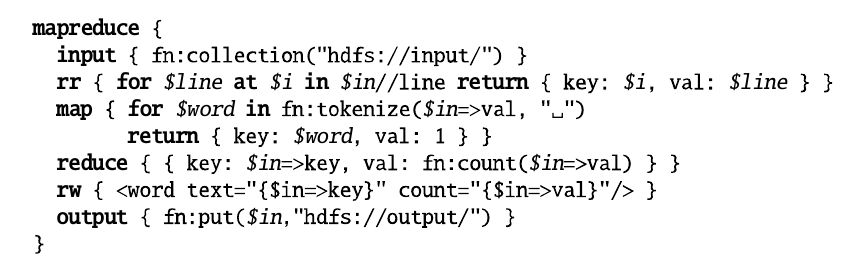}\\
  \caption{The Word Count Example Program in ChuQL~\cite{XMLMR2}}\label{Fig:CHUQL}
\end{figure}

XML (e\emph{X}tensible \emph{M}arkup \emph{L}anguage)~\cite{XMLSpecs} has been acknowledged as the defacto standard for data representation and exchange over the World Wide Web. It has found practical application in numerous domains including data interchange, streaming data and data storage. In addition, it has been considered as a standard format for many industries such as    government\footnote{http://www.irs.gov/efile/}, finance\footnote{http://www.fpml.org/}, electronic business\footnote{http://ebxml.org./} and science\footnote{http://www.w3.org/TR/2010/REC-MathML3-20101021/}. However, so far,
the problem of large XML processing using the MapReduce framework has not been extensively considered in the research literature. Some preliminary works have been presented to tackle this problem. For example,
~\cite{XMLMR} have proposed an SQL-like query language for large-scale analysis of XML data on a MapReduce platform, called \emph{MRQL} (the \emph{M}ap-\emph{R}educe \emph{Q}uery \emph{L}anguage). The evaluation system of MRQL leverages the relational query optimization techniques and compiles MRQL queries to an algebra which is then translated to physical plans using cost-based optimizations. In particular, the query plans are represented trees that are evaluated using a plan interpreter where
each physical operator is implemented with a single MapReduce job which is parameterized by the functional parameters of the physical operator.  The data fragmentation technique of MRQL is built on top of the general Hadoop XML input format which is based on a single XML tag name. Hence, given a data split of an XML document, Hadoop's input format allows reading the document as a stream of string fragments, so that each string will contain a single complete element that has the requested XML tag name.
\emph{ChuQL}~\cite{XMLMR2} is another language that have been proposed to support distributed XML processing using the MapReduce framework. It presents a MapReduce-based extension for the syntax, grammar and semantics of \emph{XQuery}~\cite{XQuery}, the standard W3C language for querying XML documents. In particular, the ChuQL implementation takes care of distributing the computation to multiple XQuery engines running in Hadoop nodes, as described by one or more ChuQL MapReduce expressions. Figure~\ref{Fig:CHUQL} illustrates the representation of the \emph{word count} example program in the ChuQL language using its extended expressions where  the \emph{mapreduce} expression is used to describe a MapReduce job.
The  \emph{input} and \emph{output} clauses are respectively used to read and write onto HDFS.
The \emph{rr} and \emph{rw}  clauses are respectively used for describing  the record reader and writer.
The \emph{map} and \emph{reduce} clauses represent the standard map and reduce phases of the framework where they process XML values or key/value pairs of XML values to match the MapReduce model which are specified
using XQuery expressions.

\subsection{MapReduce for Large Scale RDF Processing}
\label{SEC:RDFMAPREDUCE}
 RDF (\emph{R}esource \emph{D}escription \emph{F}ramework) is the standard W3C recommendation for expressing and exchanging semantic metadata about remote information sources~\cite{RDF}.
 It represent a core component of the Semantic Web initiatives as it defines a model for describing relationships among resources in terms of uniquely identified attributes and values. The basic building block in RDF is a simple tuple model, ($subject, predicate, object$), to express different types of knowledge in the form of fact statements. The interpretation of each statement is that subject $S$ has property $P$ with value $O$, where $S$ and $P$ are resource URIs and $O$ is either a URI or a literal value.  Thus, any object from one triple can play the role of a subject in another triple which amounts to chaining two labeled edges in a graph-based structure. The SPARQL query language is the official W3C standard for querying and extracting information from RDF graphs~\cite{SPARQL}. It is based on a powerful graph matching facility which allows binding variables to components in the input RDF graph and supports conjunctions and disjunctions of triple patterns.

Some research efforts have been proposed for achieving scalable RDF processing using the MapReduce framework.
\emph{PigSPARQL}~\cite{PigSPARQL} is a system that have been introduced  to process SPARQL queries using the MapReduce
framework by translating them into \emph{Pig Latin} programs where each Pig Latin program is executed by a series of
MapReduce jobs on a Hadoop cluster. ~\cite{MDAC2010RDF} have presented a preliminary algorithm for SPARQL graph pattern matching by adopting the traditional multi-way join of the RDF triples and selecting a good join-key to avoid unnecessary iterations.
~\cite{RDFCLoud} have described a storage scheme for RDF data using HDFS where the input data are partitioned into multiple
files using two main steps: 1) The \emph{Predicate Split}  which partitions the RDF triples according to their predicates. 2)
The \emph{Predicate Object Split} (POS) which uses the explicit type information  in the RDF triples to denote that a resource is an instance of a specific class while  the remaining predicate files are partitioned according to the type of their objects.
Using summary statistics for estimating the selectivities of join operations, the authors proposed an algorithm that generates a query plan whose cost is bounded by the log of the total number of variables in the given SPARQL query. An approach for optimizing RDF graph pattern matching by reinterpreting certain join tree structures as grouping operations have been presented in~\cite{TripleAlgebra,RDFAlgebra}. The proposed approach represents the intermediate results  as sets of groups of triples called \emph{TripleGroups} and uses \emph{Nested TripleGroup Algebra} for  manipulating them.~\cite{HadoopDBDemo}  have demonstrated an approach for  storing and querying RDF data using the \emph{HadoopDB} system in conjunction with a column-oriented database~\cite{SWStore} that can provide a promising solution for supporting efficient and scalable semantic web applications. A similar approach have been presented in  ~\cite{ScalableRDF} where it replaced the column-oriented back-end database with the state-of-the-art of RDF query processors, \emph{RDF-3X}~\cite{RDF3X}.

\subsection{MapReduce for Large Scale Graph Processing}
\label{SEC:GRAPHMAPREDUCE}
Graphs are popular data structures which are used to model structural relationship between objects.
Recently, they have been receiving increasing research attention as they are used in a wide variety of high impact applications  such as  social networks,  computer networks, telecommunication networks, recommendation systems, protein-protein interaction networks and the World Wide Web. Some research efforts have been proposed for providing scalable processing mechanisms for massive graph datasets.  For example, \emph{Surfer}~\cite{GraphCloud} is a large scale graph processing engine which is designed to provide two basic primitives for programmers: \emph{MapReduce} and \emph{propagation}. In this engine, MapReduce processes different key-value pairs in parallel, and propagation is an iterative computational pattern that transfers information along the edges from a vertex to its neighbors in the graph. In principle, these two primitives are complementary in graph processing where MapReduce is suitable for processing flat data structures (e.g. vertex-oriented tasks) while propagation is optimized for edge-oriented tasks on partitioned graphs.~\cite{SPAA2011} presented a set of MapReduce-based algorithms for a variety of fundamental graph problems such as minimum spanning trees, maximal matchings, approximate weighted matchings, approximate vertex and edge covers, and minimum cuts. All of the  presented algorithms are parameterized by the amount of memory available on the machines which are used to determine the number of MapReduce rounds.

\emph{GBASE}~\cite{GBASE} has been introduced as a scalable and general graph management system.
It uses graph storage method, called \emph{block compression}, to efficiently store homogeneous regions of graphs.
In particular, given the original raw graph which is stored as a big edge file, GBASE first partitions it into several homogeneous blocks. According to the partition results, it reshuffles the nodes so that the nodes belonging to the same partition are put nearby after which it compresses all non-empty block through standard compression such as \emph{GZip}\footnote{http://www.gzip.org/}. Finally, it stores the
compressed blocks together with some meta information into the graph databases.
GBASE supports different types of graph queries including \emph{neighborhood}, \emph{induced subgraph}, \emph{egonet}, \emph{K-core} and \emph{cross-edges}.
To achieve this goal, it applies a grid selection strategy to minimize disk accesses and answer queries  by
applying a MapReduce-based algorithm that supports incidence matrix based queries.
The key of query execution engine is that it unifies the different types of inputs as query vectors and unifies the different types of operations on the graph by a unified matrix-vector multiplication.
Hence,  GBASE handles queries by executing appropriate block matrix-vector multiplication modules.
\emph{PEGASUS}~\cite{PEGASUS,PEGASUS2} is another system which has been introduced as a large scale graph mining library that is implemented on the top of Hadoop and supports performing typical graph mining tasks such as \emph{computing the diameter of the graph}, \emph{computing the radius of each node} and \emph{finding the connected components} via a generalization of matrix-vector multiplication
(\emph{GIM-V}).~\cite{BiliionGraph} have presented a MapReduce-based algorithm for discovering patterns on near-cliques and triangles on large scale graphs which is built on the top of Hadoop.

In general, graph algorithms can be written as a series of chained MapReduce invocations that requires passing the entire state of the graph from one stage to the next. However, this approach is ill-suited for graph processing and can lead to suboptimal
performance due to the additional communication and associated serialization overhead in addition to the need of coordinating the steps of a chained MapReduce. The \emph{Pregel} system~\cite{Pregel} has been introduced by Google as scalable platform for implementing graph algorithms. It relies on a vertex-centric approach, which is inspired by the \emph{B}ulk \emph{S}ynchronous \emph{P}arallel model (BSP)~\cite{BSP}, where programs are expressed as a sequence of iterations, in each of which a vertex can receive messages sent in the previous iteration, send messages to other vertices and modify its own state as well as  that of its outgoing edges or mutate graph topology. In particular, Pregel computations consist of a sequence of iterations, called \emph{supersteps}. During a superstep the framework invokes a user-defined function for each vertex, conceptually in parallel, which specifies the behavior at a single vertex $V$ and a single superstep $S$. It can read messages sent to $V$ in superstep $S-1$, send messages to other vertices that will be received at superstep $S + 1$, and modify the state of $V$ and its outgoing edges. Messages are typically sent along outgoing edges, but a message may be sent to any vertex whose identifier is known. Similar to the MapReduce framework, Pregel has been designed to be  efficient, scalable and fault-tolerant implementation on clusters of thousands of commodity computers where the distribution-related details are hidden behind an abstract. It keeps vertices and edges on the machine that performs computation and uses network transfers only for messages. Hence, the model is well suited for distributed implementations as it doesn't expose any mechanism for detecting order of execution within a superstep, and all communication is from superstep $S$ to
superstep $S + 1$.  The ideas of Pregel have been cloned by many open source projects such as \emph{GoldenOrb}\footnote{http://goldenorbos.org/}, \emph{Apache Hama}\footnote{http://hama.apache.org/} and \emph{Apache Giraph}\footnote{http://giraph.apache.org/}. Both of Hama and Giraph are implemented to be launched as a typical Hadoop job that can leverage the Hadoop infrastructure. Other large scale graph processing systems which have been introduced that neither follow the MapReduce model nor leverage  the Hadoop infrastructure include \emph{GRACE}~\cite{GRACE}, \emph{GraphLab}~\cite{Graphlab2,GraphLab} and \emph{Signal/Collect}~\cite{SignalCollect}.

\subsection{Other MapReduce Applications}
\label{SEC:OTHERMAPREDUCE}
Several approaches have been proposed in the literature for tackling different data-intensive and computationally expensive operations using the MapReduce framework. For example, The \emph{Dedoop} system (\emph{De}duplication with Ha\emph{doop})~\cite{Dedoop,Dedoop2} has been presented as an entity resolution framework based on MapReduce.
It supports the ability to define complex entity resolution workflows that can include different matching steps and/or apply machine learning mechanisms for the automatic generation of match classifiers. The defined workflows are then automatically translated into MapReduce jobs for parallel execution on Hadoop clusters.
The \emph{MapDupReducer}~\cite{MapDupReducer} is another system that has been proposed as a MapReduce-based solution which is developed for supporting the problem of near duplicate detection over massive datasets using the \emph{PPJoin} (\emph{P}ositional and \emph{P}refix filtering) algorithm~\cite{PPJoin}.

An approach  to efficiently perform set-similarity joins in parallel using the MapReduce framework. In particular, they propose a 3-stage approach for end-to-end set-similarity joins have been have proposed in~\cite{SJMR}. The approach takes as input a set of records and outputs a set of joined records based on a set-similarity condition. It partitions the data across nodes in order to balance the workload and minimize the need for replication.~\cite{BruteForce} has presented three MapReduce algorithms for computing pairwise similarity on document collections. The first algorithm is based on brute force, the second algorithm treats the problem as a large-scale ad hoc retrieval and the third algorithm is based on the Cartesian product of postings lists.
\emph{V-SMART-Join}~\cite{VSMART} is a MapReduce-based framework for discovering all pairs of similar entities which is applicable to sets, multisets, and vectors. It presents a family of 2-stage algorithms where the first stage computes and joins the partial results, and the second stage computes the similarity exactly for all candidate pairs.~\cite{FuzzyJoins} have provided a theoretical analysis of various MapReduce-based similarity join algorithms in terms of various parameters including map and reduce costs, number of reducers and communication cost.

The \emph{DisCo} (\emph{Dis}tributed \emph{Co}-clustering) framework~\cite{DisCo} has been introduced as an approach for distributed data pre-processing and co-clustering from the raw data to the end clusters using the MapReduce framework.~\cite{ClusteringMD} have  presented an approach for finding subspace clusters in very large moderate-to-high dimensional data that is having typically more than 5 axes.~\cite{FastClustering} described the design and the MapReduce-based implementations of the \emph{k-median} and \emph{k-center}  clustering algorithms.
\emph{PLANET} (\emph{P}arallel \emph{L}earner for \emph{A}ssembling \emph{N}umerous \emph{E}nsemble \emph{T}rees) is a  distributed framework for learning tree models over large datasets. It defines tree learning as a series of distributed computations and implements each one using the MapReduce model~\cite{PLANET}.
The \emph{SystemML}~\cite{SystemML} provides a framework for expressing machine learning algorithms using
a declarative higher-level language. The algorithms expressed in SystemML are then automatically compiled and optimized into a set of MapReduce jobs that can run on a cluster of machines.
\emph{NIMBLE}~\cite{NIMBLE} provides an infrastructure that has been specifically designed to enable the rapid implementation of parallel machine learning and data mining algorithms. The infrastructure allows its users to compose parallel machine learning algorithms using reusable (serial and parallel) building blocks that can be efficiently executed using the MapReduce framework.
\emph{Mahout}\footnote{http://mahout.apache.org/} is an Apache project which is designed with the aim of building scalable machine learning libraries using the MapReduce framework. \emph{Ricardo}~\cite{Ricardo} is presented as a scalable platform for applying sophisticated statistical methods over huge data repositories. It is designed to facilitate the \emph{trading} between \emph{R} (a famous statistical software packages\footnote{http://www.r-project.org/}) and Hadoop
where each trading partner performs the tasks that it does best. In particular, this trading is done in a way where  \emph{R } sends aggregation-processing queries to Hadoop while Hadoop sends aggregated data to \emph{R} for advanced statistical processing or visualization.

~\cite{SpatialMR} presented an approach for  applying the MapReduce model in the domain of spatial data management. In particular, they focus on the bulk-construction of R-Trees and aerial image quality computation which involves vector and raster data. ~\cite{SocialMatch} have presented two matching algorithms, \emph{GreedyMR} and \emph{StackMR}, which are geared for the MapReduce paradigm with the aim of distributing content from information suppliers to information consumers on social media applications. In particular, they  seek to maximize the overall relevance of the matched content from suppliers to consumers while regulating the overall activity.

\end{document}